\documentclass[10pt, twocolumn, comsoc]{IEEEtran}

\usepackage{graphicx}
\usepackage[noadjust]{cite}

\usepackage{amsmath,amssymb,amsfonts}
\usepackage{fancyhdr}
\usepackage{threeparttable}
\usepackage{epsf,epsfig}
\usepackage{amsthm}
\usepackage{siunitx}
\usepackage{dsfont}
\usepackage{subfigure}
\usepackage{xcolor}
\usepackage{multicol,amsmath,lipsum}
\usepackage{graphicx}
\usepackage{epstopdf}
\usepackage{stfloats}
\usepackage[linesnumbered,ruled]{algorithm2e}
\usepackage{enumerate}
\usepackage{cancel}
\usepackage{bm}
\usepackage{newtxtext,newtxmath}
\usepackage{comment}
\usepackage{lettrine}

\usepackage[hidelinks]{hyperref}

\newtheorem{corollary}{Corollary}

\newtheorem{lemma}{Lemma}

\newtheorem{remark}{Remark}

\usepackage{amsthm}

\SetKwInput{KwInput}{Initialize}
\SetKwInput{KwOutput}{Output}
\SetKwProg{Fn}{Stage 1}{}{}
\SetKwProg{Fn}{Stage 2}{}{}

\usepackage{eucal}

\begin{document}

\title{
Reducing Latency by Eliminating CSIT Feedback: FDD Downlink MIMO Transmission for Internet-of-Things Communications
}

\author{Juntaek Han, Namhyun~Kim, and Jeonghun~Park 

\thanks{This work was supported by the Institute of Information \& communications Technology Planning \& Evaluation (IITP) grant funded by the Korea government (MSIT) (No. RS-2024-00434743, YKCS Open RAN Global Collaboration Center and No. RS-2024-00428780, 6G Cloud Research and Education Open Hub), and in part by the 6GARROW project which has received funding from the Smart Networks and Services Joint Undertaking (SNS JU) under the European Union’s Horizon Europe research and innovation programme (Grant Agreement No. 101192194) and from IITP (No. RS-2024-00435652).

Juntaek Han, Namhyun Kim and Jeonghun Park are with the School of Electrical and Electronic Engineering, Yonsei University, Seoul 03722, South Korea. (E-mail: {\texttt{jthan1218, namhyun, jhpark@yonsei.ac.kr}})
}}

\IEEEoverridecommandlockouts
\IEEEpubid{\begin{minipage}{\columnwidth}\scriptsize 
    Copyright~\copyright~2025 IEEE. Personal use of this material is permitted. However, permission to use this material for any other purposes must be obtained from the IEEE by sending a request to pubs-permissions@ieee.org.
\end{minipage}\hspace{\columnsep}\makebox[\columnwidth]{ }}

\maketitle \setcounter{page}{1}

\IEEEpubidadjcol

\begin{abstract}

This paper presents a novel framework for low-latency frequency division duplex (FDD) multi-input multi-output (MIMO) transmission with Internet of Things (IoT) communications. 
Our key idea is eliminating feedback associated with downlink channel state information at the transmitter (CSIT) acquisition. Instead, we propose to reconstruct downlink CSIT from uplink reference signals by exploiting the frequency invariance property of channel parameters. 
Nonetheless, the frequency disparity between the uplink and downlink makes it impossible to get perfect downlink CSIT, resulting in substantial interference.
To address this, we formulate a max-min fairness problem and propose a rate-splitting multiple access (RSMA)-aided efficient precoding method. 
In particular, to fully harness the potential benefits of RSMA, we propose a method that approximates the error covariance matrix and incorporates it into the precoder optimization process. This approach effectively accounts for the impact of imperfect CSIT, enabling the design of a robust precoder that efficiently handles CSIT inaccuracies.
Simulation results demonstrate that our framework outperforms other baseline methods in terms of the minimum spectral efficiency when no direct CSI feedback is used.
Moreover, we show that our framework significantly reduces communication latency compared to conventional CSI feedback-based methods, underscoring its effectiveness in enhancing latency performance for IoT communications.
\end{abstract}


\IEEEpeerreviewmaketitle

\section{Introduction}

A key distinguishable characteristic of Internet of Things (IoT) communications is the demand for extremely low latency and high reliability, often referred to as Ultra-Reliable Low Latency Communications (URLLC) \cite{osse:commmag:14, durisi:proc:16}. 
For reducing communication latency, one promising approach that has been actively studied in the literature is decreasing size of blocklength. 
As the blocklength decreases, communication systems enter the finite blocklength regime, where the block error rate does not vanish, and the classical Shannon capacity no longer accurately reflects the communication performance metric. 
To address this, the finite blocklength capacity was revealed in \cite{poly:tit:10}. 
Compared to the classical Shannon capacity, the finite blocklength capacity accounts for the non-negligible block error rate. This block error rate, along with the blocklength, determines a back-off factor for the achievable rate, in which the interplay between the rate, the blocklength, and the error rate is properly captured \cite{choi:iotj:21}. 

Harnessing the finite blocklength capacity result, abundant prior work has been presented to develop efficient design for URLLC IoT communications. 
In \cite{yang:tit:14}, the multiple-input multiple-output (MIMO) capacity in the finite blocklength regime was characterized. 
In \cite{ghanem:tcom:20}, a joint resource allocation algorithm was developed exploiting semi-definite programming for maximizing the sum spectral efficiency. 
In \cite{schiessl:tcom:18, schiessl:jsac:19}, the downlink communication latency performance was analyzed using tools of network calculus. 
Later, this was extended by incorporating uplink and non-orthogonal multiple access (NOMA) in \cite{schiessl:twc:20}. 
In \cite{choi:iotj:21}, a MIMO precoding method for jointly optimizing the achievable rate and the error probability was developed. 
In a similar vein, \cite{kim:twc:22} developed a MIMO precoding design strategy for scenarios where delay-constrained IoT devices coexist with delay-tolerant devices. 
In \cite{zhu:iotj:24, wang:jsac:23, xu:tvt:22}, a rate-splitting multiple access (RSMA) strategy was studied in the finite blocklength regime. 
Considering that the gap between the classical Shannon capacity and the finite blocklength capacity is noticeable when the number of information bits is approximately less than $20,000$ \cite{yang:tit:14} (for more than $20,000$ bits, the rate back-off factor is less than $0.1$ at the signal-to-noise ratio (SNR) of $10$ dB and the error probability of $10^{-8}$ \cite{choi:iotj:21}), the prior work is useful to design IoT communications with relatively small payloads, particularly when the payload size is less than $2.5$ KB (kilobytes). 
According to \cite{kim:proc:19}, this payload size corresponds to haptic messages in teleoperations or mission-critical messages in automotive applications. 

In \cite{kim:proc:19}, relatively heavy payload IoT communications were also discussed, including immersive virtual reality or video message for automotive applications and Internet-of-Drones (IoD), wherein the payload size ranges from $2.5 - 20$ KB. In this case, decreasing blocklength is not effective, while latency is primarily determined by spectral efficiency performance and communication overhead. 
For this reason, minimizing communication overhead while maintaining high spectral efficiency is crucial for enabling URLLC for these heavy payload IoT communications. 
Especially, the amount of overhead significantly increases when it comes with downlink MIMO systems using frequency division duplex (FDD). This is because i) in downlink MIMO, channel state information at the transmitter (CSIT) is essential for enabling multiplexing \cite{park:twc:16}; and ii) unlike in time division duplex (TDD) wherein full channel reciprocity holds, in FDD, the CSIT for the uplink and downlink bands is different \cite{xie:twc:18}. Consequently, to obtain CSIT in FDD MIMO, the transmitter must first send a downlink pilot to each user, who then computes the CSI and sends it back to the transmitter. 
This CSIT acquisition process incurs significant communication latency, which poses a challenge for supporting URLLC. 

Due to this obstacle, one may be tempted to consider alternatives to FDD MIMO, such as single-user transmission or TDD MIMO. 
Nonetheless, these solutions are not suitable for IoT communications. 
For example, with single-user transmission, spatial degrees of freedom (DoF) cannot be leveraged, resulting in significant latency when serving a large number of IoT devices. 
Furthermore, FDD is more suitable for achieving low latency, as it supports simultaneous uplink and downlink communications \cite{FDD-delay}. Additionally, it is well known that FDD provides better uplink coverage by allocating favorable low-frequency bands to IoT devices with limited transmit power \cite{FDD-Low-Freq, FDD-coverage}. This is especially critical in IoT communications, where the transmit power of IoT device is significantly limited. For this reason, enabling FDD MIMO transmission without incurring significant overhead is essential for URLLC.


There exist several prior works to reduce the overhead associated with CSIT acquisition in FDD MIMO. 
In \cite{JOMP}, a distributed CSIT compression strategy was proposed by leveraging shared geometry of scatterers for each user. Another approach \cite{caire:tit:13} reduces downlink pilot overhead by extrapolating the downlink channel covariance from the uplink covariance and selecting only its most dominant eigenvectors.
More recently, deep learning-based methods have been proposed to tackle the overhead challenge. \cite{shi:WCL:18} employs deep autoencoders for direct CSI feedback compression, while works such as \cite{guo:tcom:22, Deeplearning-Ahmed} utilize a learned channel-to-channel mapping function to extrapolate CSI to different sets of antennas and frequencies.
Another deep learning framework proposed in \cite{ma:twc:25} compresses downlink CSI to reduce feedback overhead, using a decoder that learns sparse channel representations.
In \cite{Dina}, a novel approach to reconstruct downlink CSI from uplink pilot signals was proposed, allowing the CSI feedback process to be omitted. 
This is feasible thanks to the frequency-invariant characteristics between the uplink and the downlink channels. 
For instance, even though full channel reciprocity does not hold, some key channel parameters, e.g., number of paths, angle of departure (AoD), delay, and path gains, can be assumed to be identical between the uplink and the downlink bands. 
This idea has served as the foundation for numerous other studies, such as \cite{Rottenberg:2020, Zhong:2020, Deokhwan, kim:arxiv:24}. In particular, \cite{zhang:twc:18} presented measurement campaigns to support the frequency-invariant properties.

Nonetheless, the existing work primarily focused on efficient downlink CSIT acquisition, without providing a comprehensive framework that encompasses CSI acquisition, the effects of erroneous CSI, a robust multiple-access technique, and precoder optimization. 
To address this, in \cite{kim:arxiv:24}, it was shown that 
when the downlink CSI is reconstructed from the uplink pilot signals without relying on direct CSI feedback in FDD massive MIMO systems, the robust spectral efficiency gains are only achievable if the CSI reconstruction error is properly incorporated into the precoder optimization. 

In this paper, we consider an IoT communication system using FDD MIMO. 
In such a system, we formulate a max-min fairness (MMF) problem whose main aim is to maximize the minimum spectral efficiency among the IoT devices. 
By solving this, we can guarantee the worst-case latency for the IoT communications. 
For instance, in \cite{hu:tcom:23, song:iotj:24}, the similar max-min fairness problem was tackled in the context of IoT communications, focusing on optimizing resource allocation to ensure fair performance among devices. Similarly, \cite{vu:iotj:25} investigates resource management based on max-min fairness for IoT communications.
Furthermore, some other prior work also tackled the max-min fairness problem, but it typically relied on direct CSI feedback to obtain downlink CSI \cite{kim:wcl:23, joudeh:tsp:16}, resulting in significant latency overhead. 
This approach is not suitable for IoT communications, where achieving low latency is critical. 

\begin{figure*}
\centering
\subfigure[Conventional CSI-RS based transmission]{\includegraphics[width=0.49\textwidth]{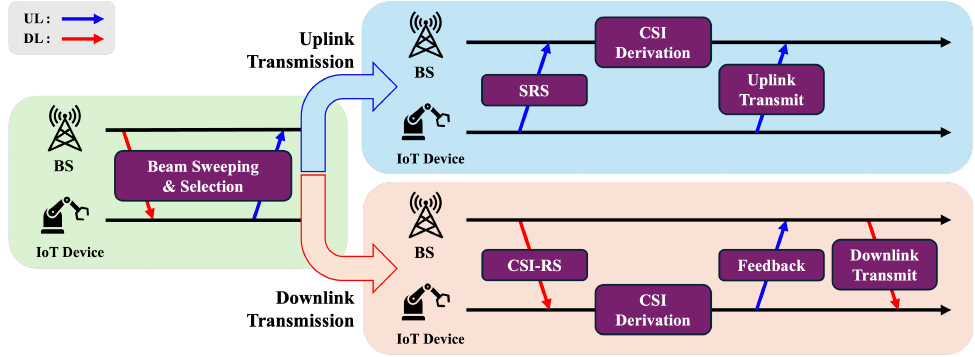}}\hfill
\subfigure[Proposed framework]{\includegraphics[width=0.49\textwidth]{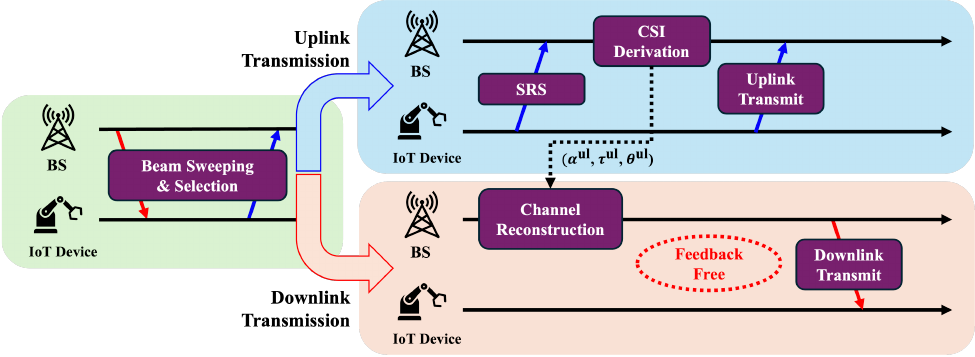}}
\caption{Comparison between the conventional CSI-RS based transmission and the proposed uplink sounding reference signal (SRS) based transmission}
\end{figure*}


To resolve the above-mentioned issue, we put forth a novel framework for enabling low-latency FDD MIMO transmission. 
Specifically, to avoid significant latency overhead associated with downlink CSIT acquisition, we reconstruct the downlink CSI from the uplink pilot signals; thereby eliminating the need for direct CSI feedback.
To this end, upon the received uplink reference signals, we use the 2D-Newtonized orthogonal matching pursuit (2D-NOMP) algorithm \cite{NOMP} to extract the key channel parameters from the uplink CSI estimation. Then, leveraging the frequency-invariant property \cite{Deokhwan, Dina, Han:2019, kim:arxiv:24}, we rebuild the downlink CSI. 
Unfortunately, this downlink CSI reconstruction approach cannot be perfect due to the uplink and downlink frequency difference, resulting in the remaining interference \cite{Deokhwan, kim:arxiv:24}. 
To mitigate this, we employ a rate-splitting multiple access (RSMA) technique \cite{Park:2023, RSMA-ten-promising, kim:wcl:23}, wherein each user's message is split into a common and a private part and then each common part is jointly encoded to make a common message. Each private part is individually encoded to make a private message. 
Based on this message construction, each user decodes the common message, by which successive interference cancellation (SIC) gains are achieved. Following this, the user decodes the private message with reduced interference. 
RSMA is known to achieve robust spectral efficiency performance in the presence of imperfect CSIT \cite{Park:2023, RSMA-ten-promising}. Reflecting its growing applicability, RSMA has demonstrated remarkable flexibility and robustness in various complex communication scenarios \cite{truong:jsac:25, kim:tvt:25}. Accordingly, we use RSMA to effectively manage the interference resulting from imperfect CSIT.

Nonetheless, it is not feasible to directly apply RSMA to our setup. 
This is because, in the precoder optimization, the CSIT error covariance is required to account for the impact of CSI inaccuracies on the spectral efficiency performance of RSMA. 
However, it is challenging to compute in our case due to the absence of CSI feedback. 
Moreover, conventional Gaussian error models introduced in \cite{sun:icc:18, le:tgcn:17} are unsuitable as they fail to capture the directional nature of non-linear estimation errors and require variance information unavailable in feedback-free setups. To deal with this, leveraging the finding that high-resolution estimation such as 2D-NOMP achieves near-CRLB performance \cite{Rottenberg:2020}, we devise a CSI error covariance approximation strategy based on observed Fisher information matrix (O-FIM), which is interpreted as a sampled version of the Fisher information matrix (FIM). 
Accordingly, O-FIM is closely connected to Cramér-Rao lower bound (CRLB), which corresponds to a lower-bound on the mean squared error (MSE) performance. Leveraging the approximated CSI error covariance, we develop a computationally efficient precoder design algorithm based on a generalized power iteration (GPI) approach \cite{Park:2023}.

The main contributions of this paper, in terms of latency, are as follows:
\begin{itemize}
    \item We propose a complete end-to-end framework that fundamentally eliminates the significant latency overhead associated with the conventional downlink CSIT acquisition process. In 5G New Radio (NR), this process involves CSI-reference signals (CSI-RS) and CSI feedback—including the precoding matrix indicator (PMI), rank indicator (RI), and channel quality indicator (CQI)—which typically takes 6 to 10 ms to complete \cite{3GPP}. Our framework bypasses this entire process by directly reconstructing the downlink CSI from uplink signals.
    \item To address the substantial interference caused by the inevitable imperfection of reconstructed CSIT, we employ a robust RSMA precoder. A key challenge is that robust precoding relies on CSIT error statistics, which are unavailable without feedback. Our contribution is a strategic ECM approximation method based on the O-FIM that makes the robust precoder design practical in a feedback-free setting.
    \item To ensure worst-case latency for all IoT devices, we formulate a max-min fairness problem. As RSMA-based optimization can be computationally intensive \cite{joudeh:tsp:16}, potentially negating latency gains, we develop a computationally efficient precoder design algorithm based on GPI and the LogSumExp approximation technique.
    \item Finally, we validate our framework’s effectiveness through a practical, holistic latency analysis. Unlike many studies that focus solely on spectral efficiency, we analyze the total communication latency within a realistic model based on 5G NR to demonstrate the framework’s substantial real-world impact.
\end{itemize}

Via simulations, we demonstrate that our framework achieves significant minimum spectral efficiency gains when no direct CSI feedback is used. Particularly, the proposed method achieves 12.5\% higher minimum spectral efficiency compared to that of the state-of-the-art RSMA precoder using the weighted minimum mean squared error (WMMSE) method \cite{joudeh:tsp:16}. 
These gains mainly come from the proposed error covariance matrix approximation technique, which suitably captures the impacts on remaining interference caused by CSI reconstruction error. 
Furthermore, we also analyze the latency performance of each method. 
We also consider the conventional CSI acquisition framework that relies on downlink training and feedback. 
In this comparison, we observe that the benefits of our method become more pronounced in terms of latency. 
Specifically, although the conventional CSI acquisition framework can obtain precise downlink CSI, it incurs approximately $6$ ms of latency, which significantly undermines the overall latency performance. 
Our framework reduces the latency associated with CSI acquisition by reconstructing it based on uplink pilots. 
Notwithstanding the fact that this approach inevitably introduces additional interference due to CSI reconstruction errors, we address this challenge by employing a robust multiple access (i.e., RSMA) and leveraging the error covariance matrix approximation technique. 
As a result, the proposed framework achieves significantly reduced latency, demonstrating its suitability for URLLC IoT communications.

The rest of this paper is organized as follows. Section II describes the system model. Section III explains the proposed downlink channel reconstruction method and characterizes the system performance. Section IV presents the error covariance matrix approximation method. Section V formulates the max-min fairness problem and presents the proposed precoder optimization algorithm. Finally, Section VI provides numerical results, and Section VII concludes the paper.

\emph{Notation}: The superscripts $(\cdot)^{\mathsf{T}}$, $(\cdot)^{\mathsf{H}}$, $(\cdot)^{-1}$, and $(\cdot)^{\dagger}$ denote the transpose, Hermitian, matrix inversion, and Moore-Penrose pseudo-inverse, respectively. We denote the $N\times N$ identity matrix by $\mathbf{I}_N$ and the trace by $\mathrm{tr}(\cdot)$. For two matrices $\mathbf{A}$ and $\mathbf{B}$ of the same size, $\mathbf{A}\circ\mathbf{B}$ denotes their Hadamard product. Given $\mathbf{A}_1,\ldots,\mathbf{A}_N\in\mathbb{C}^{K\times K}$, $\operatorname{blkdiag}[\mathbf{A}_1,\ldots,\mathbf{A}_N]$ denotes the block-diagonal matrix with diagonal blocks $\mathbf{A}_1,\ldots,\mathbf{A}_N$. For a vector $\mathbf{x}$, $\|\mathbf{x}\|$ denotes the $\ell_{2}$-norm.
We use $\mathbb{E}[\cdot]$ for the expectation operator and $\text{Re}\{\cdot\}$ for the real part of a complex quantity.

\section{System Model} \label{section:model}

We focus on IoT communications using FDD MIMO. 
We are particularly interested in a large payload case, typically ranging from $2.5$ to $20$ KB \cite{kim:proc:19}. 
In such a scenario, we explain our system model as follows. 

\subsection{Channel Model}
We consider that a base station (BS) equipped with $N$ antennas serves $K$ single-antenna devices. We denote a device set as $\CMcal{K}$. 
Based on this scenario, we first introduce the uplink channel model. Assuming that orthogonal frequency division multiplexing (OFDM) is used, the uplink channel consists of $M$ sub-carriers, each spaced by $\Delta f$. 
Then, following a widely-adopted multi-path signal model \cite{caire:tsp:17, Han:2019, han:tcom:19, zhang:twc:18}
the uplink channel for device $k \in \CMcal{K}$ on the $m$-th sub-carrier is given by
\begin{align}
    \mathbf{h}_k^{\text{ul}}[m]=\sum^{L_k^{\text{ul}}}_{\ell=1}\alpha^{\text{ul}}_{k, \ell}\mathbf{a}\left(\theta^{\text{ul}}_{k,\ell};\lambda^{\text{ul}}\right)e^{-j2\pi m\Delta f \tau^{\text{ul}}_{k, \ell}}\in \mathbb{C}^{N\times1},\label{uplink_channel_true}
\end{align}
where the number of channel paths and the complex path gain are \( L_k^{\text{ul}} \) and \( \alpha^{\text{ul}}_{k, \ell} \), respectively. The index $m \in \mathbb{Z}$ ranges from $\lfloor -M/2 \rfloor $ to $ \lfloor M/2 \rfloor - 1$. 
Assuming uniform linear array (ULA) is used at the BS, the array response vector \( \mathbf{a}\left(\theta^{\text{ul}}_{k,\ell};\lambda^{\text{ul}}\right) \) is constructed as
\begin{align}
    \mathbf{\textbf{a}}\left(\theta^{\text{ul}}_{k,\ell};\lambda^{\text{ul}}\right) = \left[1, e^{j2\pi\frac{d}{\lambda^{\text{ul}}}\text{sin}\theta^{\text{ul}}_{k,\ell}},\cdots, e^{j2\pi(N-1)\frac{d}{\lambda^{\text{ul}}}\text{sin}\theta^{\text{ul}}_{k,\ell}}\right]^{\sf T},\label{array_response_vector}
\end{align}
where $\theta^{\text{ul}}_{k,\ell}$ represents the angle-of-arrival (AoA) corresponding to $\ell$-th path in the channel for device $k$. $\lambda^{\text{ul}}$ is the uplink channel's carrier wavelength, and $d$ is the antenna spacing, which is $\lambda^{\text{ul}}/2$. 
We note that the wavelength variations over the uplink sub-carriers are relatively small, so that we assume that the AoA $\theta^{\text{ul}}_{k,\ell}$ is constant over all the sub-carriers. We clarify that this assumption was justified in \cite{Han:2019, T.Choi:2021}. 
In addition, $\tau^{\text{ul}}_{k, \ell}$ is the propagation delay for the $\ell$-th path of device $k$, which lies within $0 \leq\ \tau^{\text{ul}}_{k, \ell} \leq 1/\Delta f$. 
For generality and practicality, we assume no prior distribution on the channel parameters.

To simplify notation, we omit the sub-carrier index $m$ and replace $m\Delta f$ with generic $f^{\text{ul}}$. Consequently, the uplink channel \eqref{uplink_channel_true} is rewritten as
\begin{align}
    \mathbf{h}_k^{\text{ul}}=\sum^{L_k^{\text{ul}}}_{\ell=1}\alpha^{\text{ul}}_{k, \ell}\mathbf{a}\left(\theta^{\text{ul}}_{k,\ell};\lambda^{\text{ul}}\right)e^{-j2\pi f^{\text{ul}} \tau^{\text{ul}}_{k, \ell}}\in \mathbb{C}^{N\times1}.\label{uplink_channel_true_simplified}
\end{align}
Based on \eqref{uplink_channel_true_simplified}, we define the downlink channel model. Letting the carrier frequency difference between the uplink and downlink be $f$, the downlink channel is represented as
\begin{align}
\mathbf{h}^{\text{dl}}_k(f)=\sum^{L_k^{\text{dl}}}_{\ell=1}\alpha^{\text{dl}}_{k, \ell}\mathbf{a}\left(\theta^{\text{dl}}_{k,\ell};\lambda^{\text{dl}}\right)e^{-j2\pi  f\tau^{\text{dl}}_{k, \ell}}\in \mathbb{C}^{N\times1}\label{downlink_channel_true}.
\end{align}

Now we explain the frequency invariance property of channel parameters. As long as the propagation geometry remains unchanged between the uplink and downlink, the channel parameters are frequency-invariant, i.e., $L_k^{\text{dl}} = L_k^{\text{ul}} \triangleq L_k,$ $ \theta^{\text{dl}}_{k,\ell} = \theta^{\text{ul}}_{k,\ell}$, and $\tau^{\text{dl}}_{k, \ell} = \tau^{\text{ul}}_{k, \ell}$ as demonstrated in \cite{Deokhwan, Han:2019, zhang:twc:18, han:tcom:19}.
This frequency-invariance assumption is effective when the uplink-downlink frequency separation does not exceed approximately 10\% of the carrier frequency. This condition is reflected in the 3rd Generation Partnership Project (3GPP) specification \cite{3GPP_iot}, which specifies that path parameters remain unaltered for frequency excursions of up to 10\% of carrier frequency. Importantly, the actual duplex gaps of the 5G NR frequency range 1 (FR-1) FDD bands, where major IoT technologies like Narrowband-IoT (NB-IoT) and Long-Term Evolution for Machines (LTE-M) are deployed \cite{varsier:comm:2021, vaezi:coms:2022}, typically fall within a 4-9\% range of the carrier frequency. This confirms that, while the applicability is limited to duplex gaps within 10\% of the carrier frequency, the core assumption of our framework is valid and directly applicable to a wide range of practical IoT communication scenarios operating within these standard bands.

However, this does not mean that the uplink channel $\mathbf{h}_k^{\text{ul}}$ is same with the downlink channel $\mathbf{h}^{\text{dl}}_k(f)$. 
This is because the array response vectors $\mathbf{a}\left(\theta^{\text{ul}}_{k,\ell};\lambda^{\text{ul}}\right)$ and $\mathbf{a}\left(\theta^{\text{dl}}_{k,\ell};\lambda^{\text{dl}}\right)$ are different even with $ \theta^{\text{dl}}_{k,\ell} = \theta^{\text{ul}}_{k,\ell}$ since $\lambda^{\text {ul}} \neq \lambda^{\text {dl}}$. 
Also, the delay terms are also different, i.e., $e^{-j2\pi f^{\text{ul}} \tau^{\text{ul}}_{k, \ell}} \neq e^{-j2\pi  f\tau^{\text{dl}}_{k, \ell}}$ since the carrier frequencies are different. 
In addition to the above, the complex path gains $\alpha^{\text{dl}}_{k, \ell}$ and $\alpha^{\text{ul}}_{k, \ell}$ may be different since the path-loss is typically determined depending on the carrier frequency. 
To reflect this, we model these as 
\begin{align}
    \alpha^{\text{dl}}_{k,\ell} = \eta_{k,\ell} \alpha^{\text{ul}}_{k,\ell} + \sqrt{1-\eta_{k,\ell}^2}g, g\sim \mathcal{CN}(0, \sigma^2_{\text{path}, k}), \label{channel_gain_eta}
\end{align}
where $\eta_{k,\ell}$ indicates the correlation factor of the $\ell$-th path between the downlink and uplink path gains of device $k$. 
The correlation factor $\eta_{k,\ell}$ ranges from 0 to 1, where $\eta_{k,\ell} = 1, \forall (k,\ell)$ indicates the perfect reciprocity of the complex path gain between uplink and downlink, while $\eta_{k,\ell} = 0,\forall (k,\ell)$ indicates that the complex path gains are totally independent. 
In this sense, our model \eqref{channel_gain_eta} encompasses several previous assumptions as special cases. 
For example, the assumptions in \cite{Deokhwan, kim:twc:24} correspond to $\eta_{k,\ell} = 1, \forall (k,\ell)$ in our model.  

\subsection{RSMA Signal Model}

To mitigate the interference caused from the imperfect downlink CSIT reconstruction, we employ the 1-layer RSMA approach \cite{RSMA-ten-promising, Park:2023}. 
In this approach, the message $m_k, \forall k \in \CMcal{K}$, which is intended to device $k$, is split into a common part $m_{c,k}$ and a private part $m_{p,k}$. The common parts of all devices $m_\textit{c,1}, \cdots, m_\textit{c,K}$ are combined and jointly encoded into the common symbol $s_\textit{c}$. 
For $s_\textit{c}$, a public codebook, shared by all the devices in the considered network, is used; so that every device is able to decode $s_\textit{c}$. 
On the contrary to this, each private part $m_{p,k}$ is independently encoded into the private symbol $s_{p,k}$ using an individual codebook. 
We assume that the symbols $s_c, s_{p,k}$ are drawn independently from Gaussian codebook, i.e., $s_c, s_k \sim \mathcal{CN}(0,P)$.

For decoding, each device first decodes the common symbol first while treating the private symbols as noise. After that, the common symbol is removed using successive interference cancellation (SIC), thereafter each private symbol of device $k$ is decoded with a reduced amount of interference.

For signal transmission, we employ linear beamforming, wherein each symbol is superimposed and linearly combined with the beamforming vectors $\mathbf{F}\triangleq[\mathbf{f}_c, \mathbf{f}_1, \cdots \mathbf{f}_K] \in \mathbb{C}^{N\times(K+1)}$. 
Denoting $\mathbf{s}\triangleq[s_c, s_{p,1},\cdots,s_{p, K}]^\mathsf{T} \in \mathbb{C}^{(K+1)\times 1}$, 
the transmit signal $\mathbf{x}\in \mathbb{C}^{N\times1}$ is given by
\begin{align}
    \mathbf{x} &= \mathbf{Fs} = 
    \mathbf{f}_cs_c+\sum^{\textit{K}}_{i=1}\mathbf{f}_is_{p,i}.\label{transmit_signal}    
\end{align}
The transmit power constraint is 
${\text{tr}}(\mathbf{F}\mathbf{F}^{\sf H}) \leq 1$, ensuring that the total transmit power constraint is $P$. The received signal at device $k$ is given by
\begin{align}
    y_k=\mathbf{\textbf{h}}^{\text{dl}}_{k}(f)^{\sf H}(\mathbf{f}_cs_c+\mathbf{f}_ks_{p,k})+\sum^{K}_{i=1,i\neq k}\mathbf{\textbf{h}}^{\text{dl}}_{k}(f)^{\sf H}\mathbf{f}_is_{p,i}+z_k,\label{receive_signal_dl}
\end{align}
where $z_k \sim \mathcal{CN}(0,\sigma^2)$ is the additive white Gaussian noise (AWGN).


\section{Downlink channel Reconstruction and Performance characterization}

\subsection{Downlink Channel Reconstruction}

Recall that we do not use direct feedback to acquire downlink CSIT. 
Instead, we first estimate the uplink CSI, then extract the key channel parameters. Then, leveraging the frequency invariance property of the uplink and downlink channels, we rebuild the downlink CSI. 
To this end, we first denote a stacked uplink channel vector $\mathbf{u}(\tau, \theta)$ as
\begin{align}
    \mathbf{u}\left(\tau,\theta\right) = [\mathbf{p}_0^{\sf T}, \mathbf{p}_1^{\sf T}, \cdots ,\mathbf{p}_{M-1}^{\sf T}]^{\sf T} \in \mathbb{C}^{MN\times1},
\end{align}
where each sub-vector is
\begin{align}
    \mathbf{p}_i = \mathbf{a}\left(\theta;\lambda^{\text{ul}}\right)e^{-j2\pi\left(\lfloor-\frac{M}{2}\rfloor +i\right)\Delta f\tau}\in\mathbb{C}^{N\times 1},
\end{align}
Using this, assuming that all-ones uplink reference signal is used without loss of generality, the uplink reference signal of device $k$ across all the sub-carriers and antennas is represented as
\begin{align}
    \mathbf{y}^{\text{ul}}_k = \sum^{L_k}_{\ell=1}\alpha^{\text{ul}}_{k, \ell}\mathbf{u}(\tau_{k,\ell}^{\text{ul}},\theta_{k,\ell}^{\text{ul}})+\mathbf{w}_k\in \mathbb{C}^{MN\times 1}, \label{receive_signal_ul}
\end{align}
where $\mathbf{w}_k \sim \mathcal{CN}(0, \sigma^2_{{\text{est}}}\mathbf{I}_{MN})$ is an additive Gaussian noise in the uplink estimation phase. 

From \eqref{receive_signal_ul}, we extract the uplink channel parameters 
$\{\hat{\alpha}_{k,\ell}^{\text{ul}}, \hat{\tau}_{k,\ell}^{\text{ul}},\hat{\theta}_{k,\ell}^{\text{ul}}\}_{\forall k, \ell}$ from ${\bf{y}}_k^{{\text{ul}}}$. 
Note that this parameter extraction requires a much complicated process compared to the typical uplink channel estimation \cite{yin:jsac:2013}. For this purpose, we exploit the 2D-NOMP algorithm \cite{NOMP}.
In the 2D-NOMP algorithm, we interpret the parameter extraction problem based on \eqref{receive_signal_ul} as a spectral compressive sensing problem \cite{duarte:11}. 
Specifically, assuming a grid-based dictionary matrix 
\begin{align}
    \tilde {\bf{U}} = \begin{bmatrix}
        \mathbf{u}(\tau_{1},\theta_{1}) &  \mathbf{u}(\tau_{2},\theta_{2})& \cdots& \mathbf{u}(\tau_{Q},\theta_{Q})
    \end{bmatrix}, \label{eq:grid_dictionary}
\end{align}
we regard \eqref{receive_signal_ul} as a product between the dictionary matrix $\tilde {\bf{U}}$ and a $L_k$ sparse vector. 
Based on this viewpoint, the OMP algorithm \cite{tropp:tit:07} can be applied to find the sparse vector. 
Nonetheless, this approach is not sufficient since the true $\mathbf{u}(\tau_{k,\ell}^{\text{ul}},\theta_{k,\ell}^{\text{ul}})$ may not be on a predefined grid in $\tilde {\bf{U}}$ \eqref{eq:grid_dictionary}. 
To address this, the 2D-NOMP additionally refines the detected signal through Newton step. 
For the sake of completeness, we briefly outline the process of the 2D-NOMP algorithm in Algorithm \ref{algo:nomp}. 
For more detailed explanations regarding the 2D-NOMP algorithm, we refer readers to \cite{NOMP}.

After we obtain the estimated channel parameters $\{\hat \alpha_{k,\ell}^{\text{ul}}, \hat \tau_{k,\ell}^{\text{ul}}, \hat \theta_{k,\ell}^{\text{ul}}\}_{\forall k, \ell}$ using the 2D-NOMP algorithm, we get the downlink channel parameters as 
\begin{align}
    (\hat{\tau}_{k,\ell}^{\text{dl}}, \hat{\theta}_{k,\ell}^{\text{dl}}) &= (\hat{\tau}_{k,\ell}^{\text{ul}}, \hat{\theta}_{k,\ell}^{\text{ul}})\notag\\
    \hat{\alpha}^{\text{dl}}_{k,\ell} &= \eta_{k,\ell} \hat{\alpha}^{\text{ul}}_{k,\ell},\label{2D_NOMP_output}
\end{align}
where $\hat{\alpha}^{\text{dl}}_{k,i}$ is derived from \eqref{channel_gain_eta}. 
Finally, using $\{\hat \alpha_{k,\ell}^{\text{dl}}, \hat \tau_{k,\ell}^{\text{dl}}, \hat \theta_{k,\ell}^{\text{dl}}\}_{\forall k, \ell}$, we rebuild the downlink CSI as
\begin{align}
     \hat{\mathbf{h}}_k(f)=\sum^{L_k}_{\ell=1}\hat{\alpha}^{\text{dl}}_{k, \ell}\mathbf{a}\left(\hat{\theta}^{\text{dl}}_{k,\ell};\lambda^{\text{dl}}\right)e^{-j2\pi  f\hat{\tau}^{\text{dl}}_{k, \ell}}.\label{downlink_channel_est}
\end{align}
We exploit $\hat{\mathbf{h}}_k(f)$ to characterize the performance and design the precoders. 

\begin{algorithm}[t] 
    \caption{2D-NOMP Algorithm} \label{algo:nomp}
    \SetAlgoLined
    \KwOut{$\{\hat \alpha_{k,\ell}^{\text{ul}}, \hat \tau_{k,\ell}^{\text{ul}}, \hat \theta_{k,\ell}^{\text{ul}}\}_{\forall k, \ell}$}
    \For{$k \in \CMcal{K}$}{
    $\mathbf{r}_k \gets \mathbf{y}^{\text{ul}}_k $, $j \gets 1$ \\
    \Repeat{$\lVert\mathbf{u}^{\mathsf{H}}(\tau^{(j)}_k, \theta^{(j)}_k)$$\mathbf{r}_k\rVert^2 < \mathcal{\kappa}$}{
    \textbf{\textit{Step 1: New Detection}}
    \[
    (\tau^{(j)}_k, \theta^{(j)}_k) \gets \underset{(\tau_k, \theta_k)}{\text{arg max}}\frac{|\mathbf{u}^{\sf H}(\tau_k, \theta_k)\mathbf{r}_k|^2}{\lVert \mathbf{u}(\tau_k, \theta_k) \rVert^2},
    \]
    \begin{align}
        \alpha^{(j)}_k \gets \frac{\mathbf{u}^{\sf H}(\tau^{(j)}_k, \theta^{(j)}_k)\mathbf{r}_k}{\lVert \mathbf{u}(\tau^{(j)}_k, \theta^{(j)}_k) \rVert^2}.\tag{$\ast$}\label{2D_NOMP_alpha}
    \end{align}\\
    \textbf{\textit{Step 2: Refinement using Newton step}}\\
    Newton Step $\mathbf{s}(\alpha,\tau,\theta)$
    \[
    \mathbf{s}(\alpha,\tau,\theta)=-J''(\alpha,\tau,\theta)^{-1}J'(\alpha,\tau,\theta)
    \]
    \[
    \begin{bmatrix}
    \tau_k^{(j)} \\
    \theta_k^{(j)}
    \end{bmatrix}
    \gets
    \begin{bmatrix}
    \tau^{(j)}_k \\
    \theta^{(j)}_k
    \end{bmatrix}
    +\mathbf{s}(\alpha^{(j)}_k,\tau^{(j)}_k,\theta^{(j)}_k)
    \]\\
    Update $\alpha^{(j)}_k$ using \eqref{2D_NOMP_alpha}.\\
    Cyclical refinement using Newton step to get
    \begin{align}
        \{\alpha^{(i)}_k, \tau^{(i)}_k, \theta^{(i)}_k\}_{i=1,\cdots,j} \notag
    \end{align}\\
    \textbf{\textit{Step 3: Update gains using LS estimation}}\\
    $\mathbf{U} = [\mathbf{u}(\tau^{(1)}_k, \theta^{(1)}_k), \cdots, \mathbf{u}(\tau^{(j)}_k, \theta^{(j)}_k)]$
    \begin{align}
        [\alpha^{(1)}_k,\cdots, \alpha^{(j)}_k]^{\sf T} \gets \mathbf{U}^{\dag}\mathbf{y}_k^{\text{ul}}\notag
    \end{align}\\
    $\mathbf{r}_k \gets \mathbf{y}^{\text{ul}}_k - \sum_{1}^{j}\alpha^{(j)}_k\mathbf{u}(\tau^{(j)}_k, \theta^{(j)}_k)$, $j\gets j+1$\\
    
    }}
\end{algorithm}

\subsection{Performance Characterization}

Hereafter, we drop the notation of $f$ from $\boldsymbol{\hat{\mathbf{h}}}_{k}(f)$ for simplicity so that we denote $\hat {\bf{h}}_k$ as the downlink CSI of device $k$. 
It is worthwhile to note that the estimated downlink CSI $\hat {\bf{h}}_k$ cannot be perfect due to the inherent reconstruction error caused from the carrier frequency difference and the performance limitations of the 2D-NOMP algorithm \cite{Han:2019, T.Choi:2021}. 
Incorporating this, we present the estimated downlink CSI as follows. 
\begin{align}
    {\bf{h}}_k = \hat {\bf{h}}_k + {\bf{e}}_k, \; {\bf{e}}_k \neq {\bf{0}},
\end{align}
where ${\bf{e}}_k$ is the estimation error.  
With this, we rewrite the received signal \eqref{receive_signal_dl} as 
\begin{align}
        y_k&=\mathbf{h}^{\sf H}_k\mathbf{f}_cs_c+\sum^{K}_{\ell=1}\mathbf{h}^{\sf H}_k\mathbf{f}_\ell s_{p,\ell}+z_k\\
           &=\hat{\mathbf{h}}^{\sf H}_k\mathbf{f}_cs_c+\sum^{K}_{\ell=1}\hat{\mathbf{h}}^{\sf H}_k\mathbf{f}_\ell s_{p,\ell}+\mathbf{e}^{\sf H}_k\mathbf{f}_cs_c+ \sum^{K}_{i=1}\mathbf{e}^{\sf H}_k\mathbf{f}_is_{p,i}+z_k.\label{received_signal_dl_with_error}
\end{align}
The interference term $\sum_{i=1}^{K}\mathbf{e}_k^{\sf H}\mathbf{f}_i s_{p,i}$ in \eqref{received_signal_dl_with_error}, which arises from the CSI estimation error $\mathbf{e}_k$, is generally non-Gaussian. Directly calculating the mutual information for such a channel is often intractable, as it would require evaluating the differential entropy of the received signal, which lacks a closed-form expression. To address this, we adopt a robust approach by treating this interference term as an independent Gaussian random variable. From an information-theoretic perspective, this treatment yields a tractable worst-case lower bound on the achievable rate. This is because, for a given variance, the Gaussian distribution maximizes entropy, making it the most detrimental noise assumption and thus leading to a conservative but reliable rate expression \cite{Park:2023}. This method, based on the principles of generalized mutual information (GMI) presented in \cite{ganti:tit:00}, ensures that our subsequent precoder design is robust against the impact of non-Gaussian interference arising from practical CSI inaccuracies. By optimizing for this conservative bound, our framework is inherently prepared for the true, less-than-worst-case interference statistics. Then we reach a lower bound on the spectral efficiency of the common symbol $s_c$ achieved at device $k$ as

\begin{align}
    &R_{c,k}\notag\\
    &\geq^{(a)} \mathbb{E}_{\{\mathbf{e}_k\}}\left[\text{log}_2\left(1+\frac{|\hat{\mathbf{h}}^{\sf H}_k\mathbf{f}_c|^2}
    {
    \begin{Bmatrix}
        \sum^{K}_{\ell=1}|\hat{\mathbf{h}}^{\sf H}_k\mathbf{f}_\ell|^2+|\mathbf{e}^{\sf H}_k\mathbf{f}_c|^2 \\
        +\sum^{K}_{\ell=1}|\mathbf{e}^{\sf H}_k\mathbf{f}_\ell|^2+\frac{\sigma^2}{P}
    \end{Bmatrix}
    }\right)\right]\notag \\
    &\geq^{(b)} \text{log}_2\left(1+\frac{|\hat{\mathbf{h}}^{\sf H}_k\mathbf{f}_c|^2}
    {
    \begin{Bmatrix}
        \sum^{K}_{\ell=1}|\hat{\mathbf{h}}^{\sf H}_k\mathbf{f}_\ell|^2+\mathbf{f}^{\sf H}_{c} \mathbb{E}[\mathbf{e}_k\mathbf{e}_k^{\sf H}]\mathbf{f}_c\\
        +\sum^{K}_{\ell=1}\mathbf{f}^{\sf H}_{\ell} \mathbb{E}[\mathbf{e}_k\mathbf{e}_k^{\sf H}]\mathbf{f}_\ell+\frac{\sigma^2}{P}
    \end{Bmatrix}
    }\right)\notag\\
    &=^{(c)} \text{log}_2\left(1+\frac{|\hat{\mathbf{h}}^{\sf H}_k\mathbf{f}_c|^2}
    {
    \sum^{K}_{\ell=1}|\hat{\mathbf{h}}^{\sf H}_k\mathbf{f}_\ell|^2+\mathbf{f}^{\sf H}_{c} \bm{\Phi}_k\mathbf{f}_c+\sum^{K}_{\ell=1}\mathbf{f}^{\sf H}_{\ell} \bm{\Phi}_k\mathbf{f}_\ell+\frac{\sigma^2}{P}
    }\right) \notag\\
    &= \bar{R}_{c,k}, \label{instant_SE_common}
\end{align}
where (a) results from treating the CSI estimation error as independent Gaussian noise, (b) follows from Jensen's inequality, and (c) comes from 
\begin{align}
    \mathbb{E}[\mathbf{e}_k\mathbf{e}_k^{\sf H}]= \bm{\Phi}_k,
\end{align}
where $\bm{\Phi}_k$ is the error covariance matrix. 

After SIC, the common symbol is removed; and we also get a lower bound on the spectral efficiency of the private symbol $s_{p,k}$ at device $k$
\begin{align}
    &R_{p,k}\notag\\
    &\geq \mathbb{E}_{\{\mathbf{e}_k\}}\left[\text{log}_2\left(\frac{|\hat{\mathbf{h}}_k\mathbf{f}_k|^2}
    {
    \sum^{K}_{\ell=1, \ell\neq k}|\hat{\mathbf{h}}_k\mathbf{f}_\ell|^2+\sum^{K}_{\ell=1}|\mathbf{e}^{\sf H}_k\mathbf{f}_\ell|^2+\frac{\sigma^2}{P}
    }\right)\right]\notag\\
    &\geq \text{log}_2\left(1+\frac{|\hat{\mathbf{h}}_k\mathbf{f}_k|^2}{\sum^{K}_{\ell=1, \ell \neq k}|\hat{\mathbf{h}}_k\mathbf{f}_\ell|^2+\sum^{K}_{\ell=1}\mathbf{f}^{\sf H}_{\ell} \mathbb{E}[\mathbf{e}_k\mathbf{e}_k^{\sf H}]\mathbf{f}_\ell+\frac{\sigma^2}{P}}\right)\notag\\
    &= \text{log}_2\left(1+\frac{|\hat{\mathbf{h}}_k\mathbf{f}_k|^2}{\sum^{K}_{\ell=1, \ell \neq k}|\hat{\mathbf{h}}_k\mathbf{f}_\ell|^2+\sum^{K}_{\ell=1}\mathbf{f}^{\sf H}_{\ell} \bm{\Phi}_k\mathbf{f}_\ell+\frac{\sigma^2}{P}}\right)\notag\\
    &=\bar{R}_{p,k}. \label{instant_SE_private}
\end{align}
Note that the derivation is similar to \eqref{instant_SE_common}.

\begin{remark} \normalfont 
    Note that we use the Shannon capacity as our main performance metric. 
    This is reasonable because we mainly consider relatively heavy payload IoT communication scenarios, for example immersive virtual reality or video messages for automotive applications and Internet-of-Drones (IoD). In \cite{kim:proc:19}, the payload sizes of these IoT services are typically $2.5 - 20$ KB. 
    Conservatively, the rate back-off factor in this case is less than $0.1$ at the SNR of $10$ dB and the error probability of $10^{-8}$ as mentioned in the earlier section. 
    Accordingly, the rate back-off and the error probability are negligible in this regime, which validates our choice of performance metric. 
    %
    Further, we also note that the classical Shannon capacity has been widely used for IoT scenarios where latency matters \cite{mao:tcom:21, ghosh:tccn:24, qiu:iotj:24, zhang:iotj:20}. 
    
    When considering the finite blocklength regime, SIC in RSMA may not be perfect, which incurs additional interference when decoding the private message \cite{xu:tvt:22, wang:jsac:23}. This is due to the fact that in the finite blocklength regime, a non-negligible decoding error probability inevitably persists, potentially causing error propagation during the SIC process.
    Incorporating this effect into our framework is interesting future work. 

\end{remark}

\section{Error Covariance Matrix Approximation}

To evaluate $\bar R_{c,k}$ \eqref{instant_SE_common} and $\bar R_{p,k}$ \eqref{instant_SE_private}, the error covariance matrix ${\bf{\Phi}}_k$ is necessary. However, it is not straightforward to obtain ${\bf{\Phi}}_k$ because i) the 2D-NOMP algorithm is highly non-linear and ii) no prior distribution on the channel parameters is assumed in our setup.

To address this, we exploit a concept of the FIM. Let us define the channel parameter vector as 
\begin{align}
    \boldsymbol{\psi}_k = [\boldsymbol{\psi}_{k,1}^{\sf T}, \boldsymbol{\psi}_{k,2}^{\sf T}, \cdots, \boldsymbol{\psi}_{k,L_k}^{\sf T}]^{\sf T}\in \mathbb{R}^{4L_k\times1}
\end{align}
and 
\begin{align}
    \boldsymbol{\psi}_{k,\ell}=[\tau^{\text{ul}}_{k,\ell}, \theta^{\text{ul}}_{k,\ell}, \text{Re}\{\alpha^{\text{ul}}_{k,\ell}\}, \text{Im}\{\alpha^{\text{ul}}_{k,\ell}\}]^{\sf T}\in\mathbb{R}^{4\times1},
\end{align}
where the vectors contain the true downlink channel parameters. We let the vectors $\hat{\boldsymbol{\psi}}_k$ and $\hat{\boldsymbol{\psi}}_{k,\ell}$ represent their estimated parameters. 
For now, we assume $\eta_{k, \ell} = 1, \; \forall k, \ell$ in \eqref{channel_gain_eta}
(We will relax this later). 
Then, the CRLB is derived as
\begin{align}
    \bm{\Phi}_k\succcurlyeq\textbf{\textit{C}}(f)\triangleq(\mathbf{J}_k(f))^{\sf H}\mathbf{I}^{-1}(\boldsymbol{\psi}_k)\mathbf{J}_k(f),\label{ECM}
\end{align}
where $\mathbf{J}_k(f)\in \mathbb{C}^{4L_k\times N}$ and $\mathbf{I}(\boldsymbol{\psi}_k)\in \mathbb{C}^{4L_k\times4L_k}$ denote the Jacobian and FIM, respectively. 
Specifically, the Jacobian matrix is 
\begin{align}
    \mathbf{J}_k(f) = \frac{\partial\mathbf{h}_k^{\sf T}}{\partial\boldsymbol{\psi}_k},\label{Jacobian}
\end{align}
where $\mathbf{h}_k^{\sf T}$ is from \eqref{downlink_channel_true}. 
The FIM is 
\begin{align}
    \mathbf{I}(\boldsymbol{\psi}_k) = \mathbb{E}\left[-\frac{\partial^2 \text{log} f(\mathbf{y}|\boldsymbol{\psi}_k)}{\partial\boldsymbol{\psi}_k\partial\boldsymbol{\psi}_k^{\sf T}}\right],\label{Fisher}
\end{align}
where $f(\mathbf{y}|\boldsymbol{\psi}_k)$ represents the likelihood of $\mathbf{y}$ given the true parameter vector $\boldsymbol{\psi}_k$.
The MSE between ${\bf{h}}_k$ and $\hat {\bf{h}}_k$ is lower bounded by the diagonal element of ${\bf{C}}(f)$. 

In \cite{T.Choi:2021}, it was demonstrated that the 2D-NOMP algorithm achieves near-CRLB performance in terms of the MSE. This observation provides a strong evidence that the error covariance matrix can be tightly approximated by using a notion of CRLB with ${\bf{C}}(f)$, i.e, 
\begin{align}
        \hat{\bm{\Phi}}_k\simeq\textbf{\textit{C}}(f)\circ\mathbf{I}_N.\label{ECM_est}
\end{align}
Still, however, it is infeasible to compute $\mathbf{I}(\boldsymbol{\psi}_k)$ because it relies on the true uplink channel parameters ${\boldsymbol{\psi}}_k$, which cannot be obtained in practice.

To resolve this challenge, we propose to use a notion of O-FIM. 
Denoting the observed uplink reference signal as $\mathbf{y}$, under an assumption that the channel parameters $\boldsymbol{\psi}_k$ follow the Gaussian distribution, the O-FIM is obtained by
\begin{align}
    [\hat{\mathbf{I}}(\hat{\boldsymbol{\psi}}_k)]_{i,j} 
    &= \frac{2}{\sigma^2}\text{Re}\left\{\sum_{n=1}^{N}\sum_{m=1}^{M}\left(\frac{\partial \hat{\mathbf{y}}^{*}_{n,m}}{\partial\boldsymbol{\psi}_i}\frac{\partial \hat{\mathbf{y}}_{n,m}}{\partial\boldsymbol{\psi}_j}\right)- \right.\notag\\
    & \left. \sum_{n=1}^{N}\sum_{m=1}^{M}(\mathbf{y}_{n,m}-\hat{\mathbf{y}}_{n,m})^*\frac{\partial^2\hat{\mathbf{y}}_{n,m}}{\partial\boldsymbol{\psi}_i\partial\boldsymbol{\psi}_j}\right\}\Bigg|_{\boldsymbol{\psi}_k = \hat{\boldsymbol{\psi}}_k},\label{Observed FIM}
\end{align}
where $\hat{\mathbf{y}}$ is the observed uplink reference signal conditioned on $\hat{\boldsymbol{\psi}}_k$, namely 
\begin{align}
    \hat{\mathbf{y}} \triangleq \sum^{\hat{L}_k}_{\ell=1}\hat{\alpha}^{\text{ul}}_{k,\ell}\mathbf{u}(\hat{\tau}^{\text{ul}}_{k,\ell}, \hat{\theta}^{\text{ul}}_{k,\ell})\in\mathbb{C}^{MN\times1}.
\end{align}
Using the O-FIM, we approximate the error covariance matrix as 
\begin{align}
    \hat{\bm{\Phi}}_k \triangleq \hat{\textbf{\textit{C}}}(f)=(\hat{\mathbf{J}}_k(f))^{\sf H}\hat{\mathbf{I}}^{-1}(\hat{\boldsymbol{\psi}}_k)\hat{\mathbf{J}}_k(f),\label{ECM_estimated}
\end{align}
where $\hat{\mathbf{J}}_k(f) = \frac{\partial\hat{\mathbf{h}}^{\sf T}_k(f)}{\partial\hat{\boldsymbol{\psi}}_k} \in\mathbb{C}^{4\hat{L}_k\times N}$.

Note that $\hat{\bm{\Phi}}_k$ in \eqref{ECM_estimated} assumes $\eta_{k,\ell}=1$. When $\eta_{k,\ell}\neq 1$, the approximated error covariance matrix is further addressed in the following corollary.
\begin{corollary}
    For general $\eta_{k,\ell}$, the approximated error covariance matrix is modified to
    \begin{align}
    \hat{\bm{\Phi}}_k = \frac{1}{L_k}\left(\sum_{\ell=1}^{L_k}\eta_{k,\ell}^2\right)\hat{\textbf{\textit{C}}}(f) + \frac{1}{L_k}\left(\sum_{\ell=1}^{L_k}(1-\eta_{k,\ell}^2)\right)\mathbf{I}_N.\label{ECM_reciprocity}
    \end{align}
\end{corollary}
{\color{black}{
\begin{proof}
Assuming perfect reciprocity between the uplink and downlink channels, where $\eta_{k,\ell}=1,\forall k,\ell$, the error covariance of the reconstructed downlink channel is represented as follows:
    \begin{align}
    \bm{\Phi}_k = \mathbb{E}\left[\left(\sum_{\ell=1}^{L_k}\alpha_{k,\ell}^{\text{dl}}\mathbf{u}(\tau_{k,\ell}^{\text{dl}}, \theta_{k,\ell}^{\text{dl}})-\sum_{\ell=1}^{L_k}\hat{\alpha}_{k,\ell}^{\text{dl}}\mathbf{u}(\hat{\tau}_{k,\ell}^{\text{dl}}, \hat{\theta}_{k,\ell}^{\text{dl}})\right)\right .\notag\\
    \times \left . \left(\sum_{\ell=1}^{L_k}\alpha_{k,\ell}^{\text{dl}}\mathbf{u}(\tau_{k,\ell}^{\text{dl}}, \theta_{k,\ell}^{\text{dl}})-\sum_{\ell=1}^{L_k}\hat{\alpha}_{k,\ell}^{\text{dl}}\mathbf{u}(\hat{\tau}_{k,\ell}^{\text{dl}}, \hat{\theta}_{k,\ell}^{\text{dl}})\right)^{\sf H}\right].
\end{align}
Extending to the general case incorporating \eqref{channel_gain_eta}, the error covariance can be reformulated as
\begin{align}
    &\bm{\Phi}_k = \mathbb{E}\Bigg[\Bigg(\sum_{\ell=1}^{L_k}(\eta_{k,\ell} \alpha^{\text{ul}}_{k,\ell} + \sqrt{1-\eta_{k,\ell}^2}g)\mathbf{u}(\tau_{k,\ell}^{\text{ul}}, \theta_{k,\ell}^{\text{ul}}) \notag\\
    &\qquad\qquad\qquad\qquad - \sum_{\ell=1}^{L_k}\eta_{k,\ell}\hat{\alpha}_{k,\ell}^{\text{ul}}\mathbf{u}(\hat{\tau}_{k,\ell}^{\text{ul}}, \hat{\theta}_{k,\ell}^{\text{ul}})\Bigg) \notag\\
    &\times \Bigg(\sum_{\ell=1}^{L_k}(\eta_{k,\ell} \alpha^{\text{ul}}_{k,\ell} + \sqrt{1-\eta_{k,\ell}^2}g)\mathbf{u}(\tau_{k,\ell}^{\text{ul}}, \theta_{k,\ell}^{\text{ul}}) \notag\\
    &\qquad\qquad\qquad\qquad - \sum_{\ell=1}^{L_k}\eta_{k,\ell}\hat{\alpha}_{k,\ell}^{\text{ul}}\mathbf{u}(\hat{\tau}_{k,\ell}^{\text{ul}}, \hat{\theta}_{k,\ell}^{\text{ul}})\Bigg)^{\sf H}\Bigg], \label{ECM_proof}
\end{align}
where $\hat{\alpha}_{k,\ell}^{\text{dl}}=\eta_{k,\ell}\hat{\alpha}_{k,\ell}^{\text{ul}}$ based on the correlation knowledge. Furthermore, there is no error correlation between different devices or paths, which means
\begin{align}
    &\mathbb{E}\left[\left(\alpha_{k,\ell}^{\text{ul}}\mathbf{u}(\tau_{k,\ell}^{\text{ul}}, \theta_{k,\ell}^{\text{ul}})-\hat{\alpha}_{k,\ell}^{\text{ul}}\mathbf{u}(\hat{\tau}_{k,\ell}^{\text{ul}}, \hat{\theta}_{k,\ell}^{\text{ul}})\right)\times\right . \notag\\
    &\left . \left(\alpha_{k',\ell'}^{\text{ul}}\mathbf{u}(\tau_{k',\ell'}^{\text{ul}}, \theta_{k',\ell'}^{\text{ul}})-\hat{\alpha}_{k',\ell'}^{\text{ul}}\mathbf{u}(\hat{\tau}_{k',\ell'}^{\text{ul}}, \hat{\theta}_{k',\ell'}^{\text{ul}})\right)^{\sf H}\right]\approx0,\notag\\
    &\forall k\neq k' \text{ or }\ell\neq\ell'.
\end{align}
Using the relationship between $g$ and $(\alpha_{k,\ell}^{\text{ul}},\hat{\alpha}_{k,\ell}^{\text{ul}})$, \eqref{ECM_proof} is rewritten as follows:
\begin{align}
    &\mathbb{E}\left[\sum_{\ell=1}^{L_k}\eta^2_{k,\ell}(\alpha_{k,\ell}^{\text{ul}}\mathbf{u}(\tau_{k,\ell}^{\text{ul}}, \theta_{k,\ell}^{\text{ul}})-\hat{\alpha}_{k,\ell}^{\text{ul}}\mathbf{u}(\hat{\tau}_{k,\ell}^{\text{ul}}, \hat{\theta}_{k,\ell}^{\text{ul}}))\times \right .\notag\\
    &\left . (\alpha_{k,\ell}^{\text{ul}}\mathbf{u}(\tau_{k,\ell}^{\text{ul}}, \theta_{k,\ell}^{\text{ul}})-\hat{\alpha}_{k,\ell}^{\text{ul}}\mathbf{u}(\hat{\tau}_{k,\ell}^{\text{ul}}, \hat{\theta}_{k,\ell}^{\text{ul}}))^{\sf H}+\right .\notag\\
    &\left .  \sum_{\ell=1}^{L_k} (1-\eta^2_{k,\ell})gg^{\sf H}\mathbf{u}(\tau_{k,\ell}^{\text{ul}}, \theta_{k,\ell}^{\text{ul}})\mathbf{u}(\tau_{k,\ell}^{\text{ul}}, \theta_{k,\ell}^{\text{ul}})^{\sf H}\right]\notag\\
    &=\frac{1}{L_k}\left(\sum_{\ell=1}^{L_k}\eta^2_{k,\ell}\right)\bm{\Phi}_k+\frac{1}{L_k}\left(\sum_{\ell=1}^{L_k}(1-\eta^2_{k,\ell})\right)\mathbf{I}_N,
\end{align}
where $\mathbb{E}[gg^{\sf H}]=\sigma^2_{\text{path},k}=1/(NL_k)$ by channel normalization. We get \eqref{ECM_reciprocity} by replacing $\bm{\Phi}_k$ with $\hat{\textbf{\textit{C}}}(f)$, which completes the proof.
\end{proof}
}}
We use $\hat{\bm{\Phi}}_k$ as the approximated error covariance matrix in place of ${\bm{\Phi}}_k$. 
Now we are ready to optimize the precoders.

\begin{remark} \normalfont
    The computational complexity of the proposed O-FIM-based ECM approximation is primarily driven by the O-FIM matrix inversion and associated matrix products, resulting in a per-device complexity of $\mathcal{O}(NL_k^2+L_k^3)$. Since the ECM for each of the $K$ devices is computed independently, the total complexity scales linearly as $\mathcal{O}(K(NL_k^2+L_k^3))$. While this complexity grows with $K$, we note that the practical scalability in dense deployments is fundamentally constrained by the 3GPP standard \cite{3gpp_fim} itself rather than by our algorithm. For instance, the number of orthogonal physical downlink shared channel (PDSCH) demodulation reference signal (DM-RS) ports is limited to 12, which caps the number of devices that can be spatially multiplexed in the same time-frequency resource. Therefore, the computational cost of our ECM approximation remains well within the practical limits imposed by the communication standard for simultaneously served devices. Nonetheless, designing low-complexity algorithms for computing the O-FIM could further alleviate the computational burden, representing a valuable direction for future work.
\end{remark}

\begin{remark} \normalfont
In contrast to purely data-driven approaches, our model-based framework offers distinct advantages for latency-sensitive applications. We highlight two key aspects: First, our approach avoids the need for massive, location-specific training datasets, which are challenging to collect and maintain as the propagation environment changes \cite{guo:tcom:22, Deeplearning-Ahmed}. 

Second, and more critically, our framework provides a principled method for approximating the ECM. This statistical information is essential for designing a robust precoder that can guarantee worst-case spectral efficiency in the presence of CSI errors—a central goal of this work. In contrast, deep learning methods typically provide only a point estimate of the channel, which complicates the design of such robust schemes. Thus, our model-based approach provides a more direct and practical solution for enabling robust, low-latency communications without feedback.
\end{remark}

\section{Max-Min Fairness Precoding Optimization}

At this point, we obtain the estimated downlink CSI $\hat {\bf{h}}_k$ and the approximated error covariance matrix $\hat {\bf{\Phi}}_k$. Accordingly, we are able to evaluate $\bar R_{c,k}$ \eqref{instant_SE_common} and $\bar R_{p,k}$ \eqref{instant_SE_private}. 
Building on this, we formulate the MMF problem and develop an efficient optimization method to solve the problem. 

\subsection{Problem Formulation}

In RSMA, the common rate is determined as the minimum value among $\bar{R}_{c,k}$ for $\forall k\in\CMcal{K}$, i.e, $R_{c}\triangleq \min_{k\in\CMcal{K}}(\bar{R}_{c,k})$. 
With this, device $k$'s portion included in the common rate is denoted as $C_k\geq0$, where $\sum^K_{\ell =1}C_\ell=R_{c}$. Thus, the total rate achieved by device $k$ is defined as $C_k+\bar{R}_{p,k}$. Taking this into account, the MMF problem is formulated as follows:
\begin{align}
    &\underset{{\bf{f}}_c, \{{\bf{f}}_k\}_{k\in\CMcal{K}}, \mathbf{c}}{\text{maximize}}\quad \min_{k\in\CMcal{K}}\left(  C_k + \bar{R}_{p,k}  \right) \label{objective_function_original}\\
    &\text{subject to}\quad\min_{k\in\CMcal{K}}\left(\bar{R}_{c,k}\right)\geq\sum^{K}_{\ell=1}C_{\ell},\\
    &\quad\quad\quad\quad\quad \ C_k \geq 0, \forall k \in \CMcal{K},\notag\\
    &\quad\quad\quad\quad\quad \ \text{tr}\left( \mathbf{FF}^{\sf H}   \right) \leq 1,\label{power_constraint}
\end{align}
where $\mathbf{c}=[C_1, \cdots , C_K]^\mathsf{T}\in\mathbb{C}^{K\times 1}$. 
Unfortunately, finding the global optimal solution of \eqref{objective_function_original} is challenging due to its non-convexity and non-smoothness. 
Addressing this, we reformulate \eqref{objective_function_original} into a tractable form in the next subsection. 

\subsection{Reformulation into a Tractable Form}
We first apply the LogSumExp (LSE) technique to approximate the non-smooth minimum function in \eqref{objective_function_original}. With the LSE, the minimum function is approximated as 
\begin{align}
    \text{LSE}\left\{\min_{k\in\CMcal{K}}\left(  C_k + \bar{R}_{p,k}\right)\right\}\approx -\alpha\text{log}\left( \sum^{K}_{i=1} \text{exp}\left(\frac{C_i+\bar{R}_{p,i}}{-\alpha}\right)\right)\label{LSE_objective},
\end{align}
where \eqref{LSE_objective} becomes tight as $\alpha \rightarrow 0+$. 
Now, we arrange the beamforming vectors $\textbf{f}_c, \textbf{f}_1, \cdots \textbf{f}_K$ by stacking them into a single vector denoted as $\bar{\mathbf{f}} \triangleq [\textbf{f}_c^{\mathsf{T}}, \textbf{f}_1^{\mathsf{T}}, \cdots \textbf{f}_K^{\mathsf{T}}]^\mathsf{T}\in \mathbb{C}^{N\times (K+1)}$, where $\lVert \bar{\mathbf{f}} \rVert^2=1$. 
With this, we rewrite $\bar{R}_{p,k}$ and $\bar{R}_{c,k}$ in a Rayleigh quotients form, i.e., 
\begin{align}
    \bar{R}_{p,k}=\text{log}_2\left(\frac{\bar{\mathbf{f}}^{\sf H}\mathbf{A}_k\bar{\mathbf{f}}}{\bar{\mathbf{f}}^{\sf H}\mathbf{B}_k\bar{\mathbf{f}}}\right), \bar{R}_{c,k}=\text{log}_2\left(\frac{\bar{\mathbf{f}}^{\sf H}\mathbf{C}_k\bar{\mathbf{f}}}{\bar{\mathbf{f}}^{\sf H}\mathbf{D}_k\bar{\mathbf{f}}}\right),\label{new_expression_of_rate}
\end{align}
where
\begin{align}
    \mathbf{A}_k&=\text{blkdiag}\left[  \mathbf{0}, \left( \hat{\mathbf{h}}_k\hat{\mathbf{h}}^{\sf H}_k+\hat{\bm{\Phi}}_k \right), \cdots, \left( \hat{\mathbf{h}}_k\hat{\mathbf{h}}^{\sf H}_k+\hat{\bm{\Phi}}_k \right) \right]\notag\\
    &+\frac{\sigma^2}{P}\mathbf{I}_{N(K+1)}\\
    \mathbf{B}_k&=\mathbf{A}_k - \text{blkdiag}\left[\mathbf{0}, \mathbf{0}, \cdots, \underbrace{\hat{\mathbf{h}}_k \hat{\mathbf{h}}_k^{\mathsf{H}}}_{(k+1)\text{-th matrix}}, \cdots, \mathbf{0}     \right]\\
    \mathbf{C}_k&=\text{blkdiag}\left[\left( \hat{\mathbf{h}}_k\hat{\mathbf{h}}^{\sf H}_k+\hat{\bm{\Phi}}_k \right), \cdots, \left( \hat{\mathbf{h}}_k\hat{\mathbf{h}}^{\sf H}_k+\hat{\bm{\Phi}}_k \right) \right]\notag\\
    &+\frac{\sigma^2}{P}\mathbf{I}_{N(K+1)}\\
    \mathbf{D}_k&=\mathbf{C}_k - \text{blkdiag}\left[ \hat{\mathbf{h}}_k\hat{\mathbf{h}}^{\sf H}_k, \mathbf{0}, \cdots, \mathbf{0}     \right].
\end{align}
Since we can normalize both numerator and denominator of \eqref{new_expression_of_rate} with $\| \bar {\bf{f}} \|$, we omit the power constraint in \eqref{power_constraint}. 
With the LSE approximation technique and the Rayleigh quotient reformulation, the problem \eqref{objective_function_original} is transformed to 
\begin{align}
    &\underset{\bar {\bf{f}}, \mathbf{c}}{\text{maximize}}\quad\left\{(-\alpha)\text{log}\left[ \sum^{K}_{i=1} \text{exp}\left\{-\frac{1}{\alpha}\left(C_i+\text{log}_2\left(\frac{\bar{\mathbf{f}}^{\sf H}\mathbf{A}_i\bar{\mathbf{f}}}{\bar{\mathbf{f}}^{\sf H}\mathbf{B}_i\bar{\mathbf{f}}}\right)\right)\right\}\right]\right\}\label{objective_function_revised_1}\\ 
    &\text{subject to}\quad\min_{k\in\CMcal{K}}\left(\bar{R}_{c,k}\right)\geq\sum^{K}_{\ell=1}C_\ell, \notag\\
    &\quad\quad\quad\quad\quad \ C_k \geq 0, \forall k \in \CMcal{K}. \label{subject_revised_1}
\end{align}
Now we tackle \eqref{objective_function_revised_1}. 

\subsection{Proposed MMF Optimization Method}
For \eqref{objective_function_revised_1} with \eqref{subject_revised_1}, we find local optimal $\mathbf{F}$ and $\mathbf{c}$. 
To this end, we exploit the two stage algorithm \cite{kim:wcl:23}, wherein the two stages are alternated to solve the problem iteratively.  
Specifically, in the first stage, we use the GPI-based beamforming method to find $\mathbf{F}$ given $\mathbf{c}$, and the second stage, we use a waterfilling-like method to find $\mathbf{c}$ for fixed $\mathbf{F}$.

The Lagrangian function of the problem \eqref{objective_function_revised_1} is identified by
\begin{align}
    \lambda(\bar{\mathbf{f}})&=\left(-\alpha\right)\log\left[\sum_{k=1}^{K}\exp\left\{-\frac{1}{\alpha}\left(C_{k}+\log_{2}\left(\frac{\bar{\mathbf{f}}^{\sf H}\mathbf{A}_{k}\bar{\mathbf{f}}}{\bar{\mathbf{f}}^{\sf H}\mathbf{B}_{k}\bar{\mathbf{f}}}\right)\right)\right\}\right]\notag\\
    &-\gamma\left[\sum_{\ell=1}^{K}C_{\ell}+\alpha\log\left\{\sum_{k=1}^{K}\exp\left(\log_{2}\left(\frac{\bar{\mathbf{f}}^{\sf H}\mathbf{C}_{k}\bar{\mathbf{f}}}{\bar{\mathbf{f}}^{\sf H}\mathbf{D}_{k}\bar{\mathbf{f}}}\right)^{-\frac{1}{\alpha}}\right)\right\}\right],\label{lagrangian_function}
\end{align}
where $\lambda$ is the Lagrange multiplier. 
With \eqref{lagrangian_function}, we enter the first stage as follows. 

\subsubsection{Stage 1: Beamforming Design}
The first-order optimality condition of the Lagrangian function \eqref{lagrangian_function} is obtained in the following lemma.
\begin{lemma}
    The first-order optimality condition of \eqref{lagrangian_function} is satisfied if the following holds:
    \begin{align}
        \mathbf{Y}^{-1}_{\mathsf{KKT}}(\bar{\mathbf{f}})\mathbf{X}_{\mathsf{KKT}}(\bar{\mathbf{f}})\bar{\mathbf{f}}=\lambda(\bar{\mathbf{f}})\bar{\mathbf{f}}, \label{KKT_condition}
    \end{align}
    where
    {\small
    \begin{align}
        &\mathbf{X}_{\sf KKT}(\bar{\mathbf{f}})\notag\\
        &=\sum\limits_{k=1}^{K}\left\{\frac{\exp\left(-\frac{1}{\alpha}\left(C_{k}+\bar{R}_{p,k}\right)\right)\frac{\mathbf{A}_{k}}{\bar{\mathbf{f}}^{\sf H}\mathbf{A}_{k}\bar{\mathbf{f}}}}{\sum_{\ell=1}^{K}\exp\left(-\frac{1}{\alpha}\left(C_{\ell}+\bar{R}_{p,\ell}\right)\right)}+\frac{\exp\left(-\frac{1}{\alpha}\bar{R}_{c,k}\right)\frac{\gamma\mathbf{C}_{k}}{\bar{\mathbf{f}}^{\sf H}\mathbf{C}_{k}\bar{\mathbf{f}}}}{\sum_{\ell=1}^{K}\exp\left(-\frac{1}{\alpha}\bar{R}_{c,\ell}\right)}\right\},\label{X}\\
        &\mathbf{Y}_{\sf KKT}(\bar{\mathbf{f}})\notag\\
        &=\sum\limits_{k=1}^{K}\left\{\frac{\exp\left(-\frac{1}{\alpha}\left(C_{k}+\bar{R}_{p,k}\right)\right)\frac{\mathbf{B}_{k}}{\bar{\mathbf{f}}^{\sf H}\mathbf{B}_{k}\bar{\mathbf{f}}}}{\sum_{\ell=1}^{K}\exp\left(-\frac{1}{\alpha}\left(C_{\ell}+\bar{R}_{p,\ell}\right)\right)}+\frac{\exp\left(-\frac{1}{\alpha}\bar{R}_{c,k}\right)\frac{\gamma\mathbf{D}_{k}}{\bar{\mathbf{f}}^{\sf H}\mathbf{D}_{k}\bar{\mathbf{f}}}}{\sum_{\ell=1}^{K}\exp\left(-\frac{1}{\alpha}\bar{R}_{c,\ell}\right)}\right\}.\label{Y}
    \end{align}
    }
\end{lemma}

\begin{proof}
    The first-order optimality condition of \eqref{lagrangian_function} is derived as follows:
\begin{align}
    \frac{\partial \lambda(\bar{\mathbf{f}})}{\partial \bar{\mathbf{f}}^{\sf H}}&=\sum\limits_{k=1}^{K}\left\{\frac{\exp\left(-\frac{1}{\alpha}\left(C_{k}+\bar{R}_{p,k}\right)\right)\left(\frac{\mathbf{A}_{k}\bar{\mathbf{f}}}{\bar{\mathbf{f}}^{\sf H}\mathbf{A}_{k}\bar{\mathbf{f}}} - \frac{\mathbf{B}_{k}\bar{\mathbf{f}}}{\bar{\mathbf{f}}^{\sf H}\mathbf{B}_{k}\bar{\mathbf{f}}}\right)}{\sum_{\ell=1}^{K}\exp\left(-\frac{1}{\alpha}\left(C_{\ell}+\bar{R}_{p,\ell}\right)\right)}\right\}\notag\\
    &+\sum\limits_{k=1}^{K}\left\{\frac{\exp\left(-\frac{1}{\alpha}\bar{R}_{c,k}\right)\left(\frac{\gamma\mathbf{C}_{k}\bar{\mathbf{f}}}{\bar{\mathbf{f}}^{\sf H}\mathbf{C}_{k}\bar{\mathbf{f}}} - \frac{\gamma\mathbf{D}_{k}\bar{\mathbf{f}}}{\bar{\mathbf{f}}^{\sf H}\mathbf{D}_{k}\bar{\mathbf{f}}}\right)}{\sum_{\ell=1}^{K}\exp\left(-\frac{1}{\alpha}\bar{R}_{c,\ell}\right)}\right\}.\label{KKT_proof_1}
\end{align}
When \eqref{KKT_proof_1} equals to 0, the optimality condition is satisfied:
{ \small
\begin{align}
    &\sum\limits_{k=1}^{K}\left\{\frac{\exp\left(-\frac{1}{\alpha}\left(C_{k}+\bar{R}_{p,k}\right)\right)\frac{\mathbf{A}_{k}\bar{\mathbf{f}}}{\bar{\mathbf{f}}^{\sf H}\mathbf{A}_{k}\bar{\mathbf{f}}}}{\sum_{\ell=1}^{K}\exp\left(-\frac{1}{\alpha}\left(C_{\ell}+\bar{R}_{p,\ell}\right)\right)}+\frac{\exp\left(-\frac{1}{\alpha}\bar{R}_{c,k}\right)\frac{\gamma\mathbf{C}_{k}\bar{\mathbf{f}}}{\bar{\mathbf{f}}^{\sf H}\mathbf{C}_{k}\bar{\mathbf{f}}}}{\sum_{\ell=1}^{K}\exp\left(-\frac{1}{\alpha}\bar{R}_{c,\ell}\right)}\right\}\notag\\
    &=\sum\limits_{k=1}^{K}\left\{\frac{\exp\left(-\frac{1}{\alpha}\left(C_{k}+\bar{R}_{p,k}\right)\right)\frac{\mathbf{B}_{k}\bar{\mathbf{f}}}{\bar{\mathbf{f}}^{\sf H}\mathbf{B}_{k}\bar{\mathbf{f}}}}{\sum_{\ell=1}^{K}\exp\left(-\frac{1}{\alpha}\left(C_{\ell}+\bar{R}_{p,\ell}\right)\right)}+\frac{\exp\left(-\frac{1}{\alpha}\bar{R}_{c,k}\right)\frac{\gamma\mathbf{D}_{k}\bar{\mathbf{f}}}{\bar{\mathbf{f}}^{\sf H}\mathbf{D}_{k}\bar{\mathbf{f}}}}{\sum_{\ell=1}^{K}\exp\left(-\frac{1}{\alpha}\bar{R}_{c,\ell}\right)}\right\}.\label{KKT_proof_2}
\end{align}
}
Substituting \eqref{lagrangian_function}, \eqref{X}, and \eqref{Y} into \eqref{KKT_proof_2}, we get the following equation
\begin{align}
    \mathbf{X}_{\sf KKT}(\bar{\mathbf{f}})\bar{\mathbf{f}} = \lambda(\bar{\mathbf{f}})\mathbf{Y}_{\sf KKT}(\bar{\mathbf{f}})\bar{\mathbf{f}},
\end{align}
which leads to the equation \eqref{KKT_condition}.
\end{proof} 
Note that \eqref{KKT_condition} is interpreted as a non-linear eigenvector-dependent eigenvalue problem, where $\bar{\mathbf{f}}$ represents a eigenvector and $\lambda(\bar{\mathbf{f}})$, the Lagrangian function in \eqref{lagrangian_function}, is the corresponding eigenvalue. Accordingly, if we find a leading eigenvector of the matrix $\mathbf{Y}^{-1}_{\mathsf{KKT}}(\bar{\mathbf{f}})\mathbf{X}_{\mathsf{KKT}}(\bar{\mathbf{f}})$ in \eqref{KKT_condition} given the optimal Lagrange multiplier $\gamma$, it maximizes the objective function in \eqref{objective_function_revised_1} while satisfying the first-order optimality condition (its gradient is zero). 
However, it is not trivial to find a leading eigenvector of \eqref{KKT_condition} due to its dependency on the eigenvector itself. In particular, the matrix $\mathbf{Y}^{-1}_{\mathsf{KKT}}(\bar{\mathbf{f}})\mathbf{X}_{\mathsf{KKT}}(\bar{\mathbf{f}})$ in \eqref{KKT_condition} is a function of $\bar {\bf{f}}$, making classical methods such as power iteration or eigenvalue decomposition unsuitable for our problem. 
We note that if the matrix $\mathbf{Y}^{-1}_{\mathsf{KKT}}(\bar{\mathbf{f}})\mathbf{X}_{\mathsf{KKT}}(\bar{\mathbf{f}})$ does not depend on $\bar {\bf{f}}$, our problem reduces to a classical eigenvalue problem.

To deal with this, we devise a GPI-based method that iteratively finds a leading eigenvector of \eqref{KKT_condition}. Specifically, we update the beamforming vector in $t$-th iteration, $\bar{\mathbf{f}}^{(t)}$ as
\begin{align}
    \bar{\mathbf{f}}^{(t)}=\frac{\mathbf{Y}_{\mathsf{KKT}}^{-1}(\bar{\mathbf{f}}^{(t-1)})\mathbf{X}_{\mathsf{KKT}}(\bar{\mathbf{f}}^{(t-1)})\bar{\mathbf{f}}^{(t-1)}}{\lVert \mathbf{Y}_{\mathsf{KKT}}^{-1}(\bar{\mathbf{f}}^{(t-1)})\mathbf{X}_{\mathsf{KKT}}(\bar{\mathbf{f}}^{(t-1)})\bar{\mathbf{f}}^{(t-1)}\rVert}.\label{GPI_update}
\end{align}
It repeats until $\lVert \bar{\mathbf{f}}^{(t)} - \bar{\mathbf{f}}^{(t-1)}\rVert$ is bounded by the predetermined parameter $\epsilon$.\\

\subsubsection{Stage 2: Common Rate Portions Allocation}
The main purpose of the second stage is to determine $C_k$ so as to maximize the worst spectral efficiency for given $\bar{\mathbf{f}}$. 
To this end, based on the obtained $\bar{\mathbf{f}}$ in the first stage, we use a waterfilling-like method to allocate $C_k$. Here, we represent the common and private rate given $\bar{\mathbf{f}}$ as $\hat{R}_{c,k}$ and $\hat{R}_{p,k}$, where they are considered constants at this stage. 

Following the well-known waterfilling principle, we reallocate $C_k$ by 
\begin{align}
    C_k = \left(\mu - \hat R_{p,k} \right)^{+},\label{waterfilling_1}
\end{align}
where $\mu$ is determined by 
\begin{align}
    \sum_{k = 1}^{K} \left(\mu - \hat R_{p,k} \right)^{+} = \text{min}_k(\hat{R}_{c,k}).\label{waterfilling_2}
\end{align}
Applying this, more $C_k$ is allocated to devices with lower private rates, prioritizing those requiring more common rate portions to maximize the minimum rate. The term $\left(\mu - \hat R_{p,k} \right)^{+}$ determines $C_k$, with $\mu$ representing the water-level, and $x^{+} \triangleq \text{max}(x,0)$ ensures non-negative allocations. The total allocation satisfies \eqref{waterfilling_2}, where $\mu$ is adjusted so that the sum of $C_k$ matches $\text{min}_k(\hat{R}_{c,k})$.

The above two stages are repeated by gradually increasing the Lagrange multiplier $\gamma$. 
Upon completing the process, we select the best $\mathbf{\bar{f}}$ and $\mathbf{c}$ that yield the maximum objective value.

\subsection{Complexity Analysis}
The computational complexity of the proposed MMF optimization method is primarily driven by the GPI-based beamforming design in \textit{Stage 1}. The main computational cost within each GPI iteration \eqref{GPI_update} is the inversion of the block-diagonal matrix $\mathbf{Y}_{\text{KKT}}(\bar{\mathbf{f}})$. Since this matrix consists of $(K+1)$ sub-blocks of size $N \times N$, the complexity of this operation is $\mathcal{O}(KN^3)$. This entire two-stage process is repeated in an outer loop that searches for the optimal Lagrange multiplier $\gamma$. Consequently, the overall worst-case complexity of the proposed algorithm is $\mathcal{O}((\#\gamma)\log(1/\epsilon)KN^{3})$, where $\#\gamma$ is the number of outer-loop updates for $\gamma$, and the solution accuracy parameter $\epsilon$ contributes the factor $\mathcal{O}(\log(1/\epsilon))$ for convergence.

\section{Numerical Results}
In this section, we demonstrate the performances of the proposed framework. 
For the used simulations environments, we basically follow the setup described in Section \ref{section:model}. 
More specifically, we assume $N = 12, K = 4, L_k = 5 \ \text{for} \ k \in \CMcal{K}, 0.5 \leq \gamma \leq 0.9 , \sigma^2 = 1$. 
For the proposed algorithm parameters, we set $\alpha = 0.1 \text{ and }\epsilon = 0.01$. 
For the downlink channel reconstruction, we assume that $\eta_{k,\ell}=0.9\ \forall(k,\ell)$ in \eqref{channel_gain_eta} and the angular spread is $\pi/10$. The uplink SNR is fixed at 10 dB.


\begin{table*}[t] 
\centering
\renewcommand{\arraystretch}{1.3}
\caption{List of Abbreviations for Curves in Simulation Figures}
\label{tab:abbr}
\begin{tabular}{|m{2.5cm}|m{5.5cm}|m{7cm}|}
\hline\hline
\textbf{Abbreviation} & \textbf{Full name} & \textbf{Description} \\ \hline
Proposed & Proposed RSMA precoder for max-min fairness & GPI-based RSMA precoder with ECM approximation \\ \hline
Proposed (w/o ECM) & Proposed method without ECM approximation & Same as the proposed method but assuming the ECM $\hat{\boldsymbol{\Phi}}_k = \mathbf{0}$ \\ \hline
WMMSE & Weighted Minimum Mean-Squared Error & RSMA precoder optimized via the WMMSE algorithm~\cite{joudeh:tsp:16} \\ \hline
GPI (w/o RS) & Generalized Power Iteration without RSMA & GPI-based precoder but without applying RSMA \\ \hline
RZF & Regularized Zero-Forcing & Linear precoder mitigating inter-user interference via regularized matrix inversion \\ \hline
MRT & Maximum Ratio Transmission & Linear precoder aligning each beamforming vector with the corresponding estimated channel vector \\ \hline
GPI with feedback & Generalized Power Iteration with CSI feedback & GPI-based RSMA precoder assuming perfect CSI is obtained via conventional CSI feedback, i.e., $\hat{\mathbf{h}}_k=\mathbf{h}_k$  \\ \hline\hline
\end{tabular}
\end{table*}


We consider the following baseline schemes and their associated computational complexities:



\begin{itemize}
    \item \textbf{Maximum Ratio Transmission (MRT)}: The precoding vectors are aligned with the estimated channel vectors, i.e., $\mathbf{f}_k=\hat{\mathbf{h}}_k$. Its complexity is dictated by channel-vector scaling and normalization and grows linearly with the numbers of IoT devices and antennas, i.e., $\mathcal{O}(KN)$.

    \item \textbf{Regularized Zero-Forcing (RZF)}: This method eliminates the interference by $\mathbf{f}_k = (\hat{\mathbf{H}}\hat{\mathbf{H}}^{\sf H}+\frac{\sigma^2}{P}\mathbf{I})^{-1}\hat{\mathbf{h}}_k$, where $\hat{\mathbf{H}}=[\hat{\mathbf{h}}_1,\cdots, \hat{\mathbf{h}}_k]$. The matrix inversion dominates the complexity at $\mathcal{O}(K^{3})$, followed by the multiplication cost of $\mathcal{O}(K^{2}N)$, leading to an overall complexity of $\mathcal{O}(K^{3}+K^{2}N)$.

    \item {\textbf{WMMSE}}: This method uses the WMMSE algorithm to solve the MMF problem. As shown in \cite{joudeh:tsp:16}, it has a worst-case computational complexity of $\mathcal{O}(\log(1/\epsilon)(KN)^{3.5})$, where $\epsilon$ is the solution-accuracy parameter.

    \item {\textbf{GPI without RS}}: This method uses GPI as a precoding optimizer without considering RSMA. As a simplified version of our proposed algorithm, its complexity is similarly dominated by the iterative matrix operations, resulting in a worst-case complexity of $\mathcal{O}(\log(1/\epsilon)KN^{3})$, where $\epsilon$ is the solution-accuracy parameter.
\end{itemize}

We note that all the above cases adopt the reconstructed downlink CSI from the uplink pilots as in the proposed framework. The abbreviations and brief descriptions for each benchmark scheme, including the one based on CSI feedback, are summarized in Table~\ref{tab:abbr}.

\subsection{Spectral Efficiency}
\begin{figure}[t]
 \renewcommand{\figurename}{Fig.}
    \centering
    \includegraphics[width=\columnwidth]{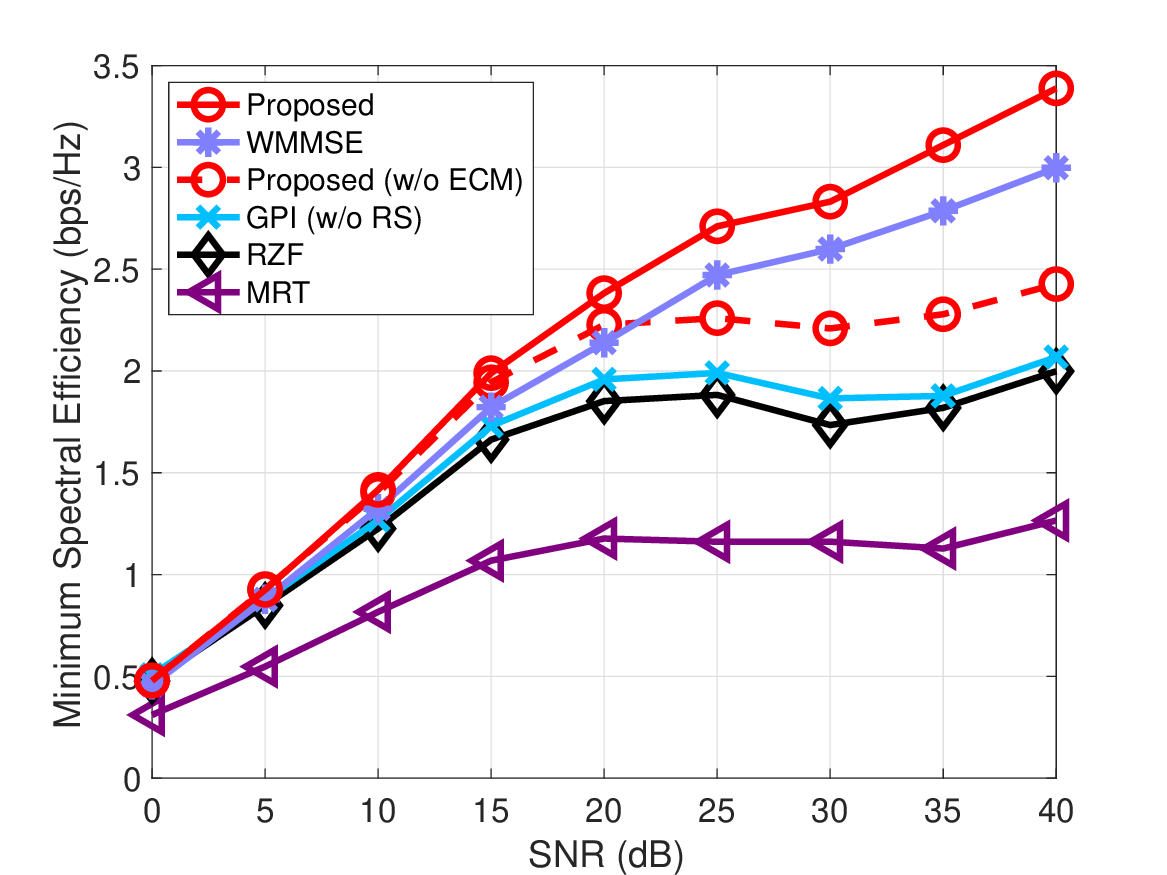} 
    \caption{Comparison of the minimum spectral efficiency among devices versus SNR for different methods.}
    \label{Fig:minimum_rate}
\end{figure}

At first, we evaluate the minimum spectral efficiency performance in Fig.~\ref{Fig:minimum_rate}.
It is worth noting that the performance gains demonstrated in this section are based on the assumption that key channel parameters remain stable during the time required to reconstruct the downlink channel from the uplink reference signals. This assumption holds for the target IoT scenarios with fixed or moderate mobility devices, as the channel coherence time is sufficiently long to accommodate the processing of uplink signals for downlink transmission. For the proposed method, we also illustrate a case that does not use the error covariance matrix approximation. In this case, we let $\hat{\bm{\Phi}}_k = \bm{0}$. 

As observed in Fig.~\ref{Fig:minimum_rate}, the proposed method outperforms other approaches, achieving performance gains of $12.54\%$ compared to WMMSE, $24.11\%$ compared to the proposed framework without error covariance matrix approximation, and $60.09\%$ compared to GPI without RS at 40 dB. We interpret this result as follows. The significant gains of the proposed method over GPI without RS highlights the advantages of RSMA under imperfect downlink CSI reconstruction.
This finding is consistent with the results in \cite{kim:wcl:23, Park:2023, RSMA-ten-promising}, demonstrating RSMA’s robustness in scenarios with imperfect CSI.

It is evident that applying RSMA without our approximation of the error covariance matrix significantly degrades the performance gains.
This is because, the interference caused by imperfect downlink CSI reconstruction is not properly accounted into the optimization, leading to degraded performance \cite{kim:arxiv:24}.  

\begin{table}[t]
\centering
\caption{Average computation time (Sec.) of RSMA precoders }
\label{Table:computation}
\begin{tabular}{c|cccc}
\noalign{\smallskip}\noalign{\smallskip}\hline\hline
$(N\times K)$ & Proposed & WMMSE \\
\hline
$(12\times 4)$ & 0.2263 & 1.9369 \\
\hline
\hline
\end{tabular}
\end{table}

We also observe the gains of the proposed method compared to WMMSE. 
As described in \cite{joudeh:tsp:16}, WMMSE leverages the rate-MSE equivalence to transform the rate maximization problem into a more tractable weighted MSE minimization problem.
In this process, the MSE-based formulation does not adequately account for the impact of interference caused by imperfect CSI, resulting in degraded performance compared to the proposed method. 
Moreover, it is also worthwhile to note that solving weighted MSE minimization problem needs to use an off-the-shelf optimization toolbox such as CVX \cite{cvx}, so that WMMSE requires an average computation time approximately 8.56 times longer than that of the proposed method (Table~\ref{Table:computation}). This adversely affects latency performance, rendering WMMSE less suitable for IoT communication scenarios and underscoring the advantages of the proposed method as a more efficient and appropriate solution.

\begin{figure}[t]
 \renewcommand{\figurename}{Fig.}
    \centering
    \includegraphics[width=\columnwidth]{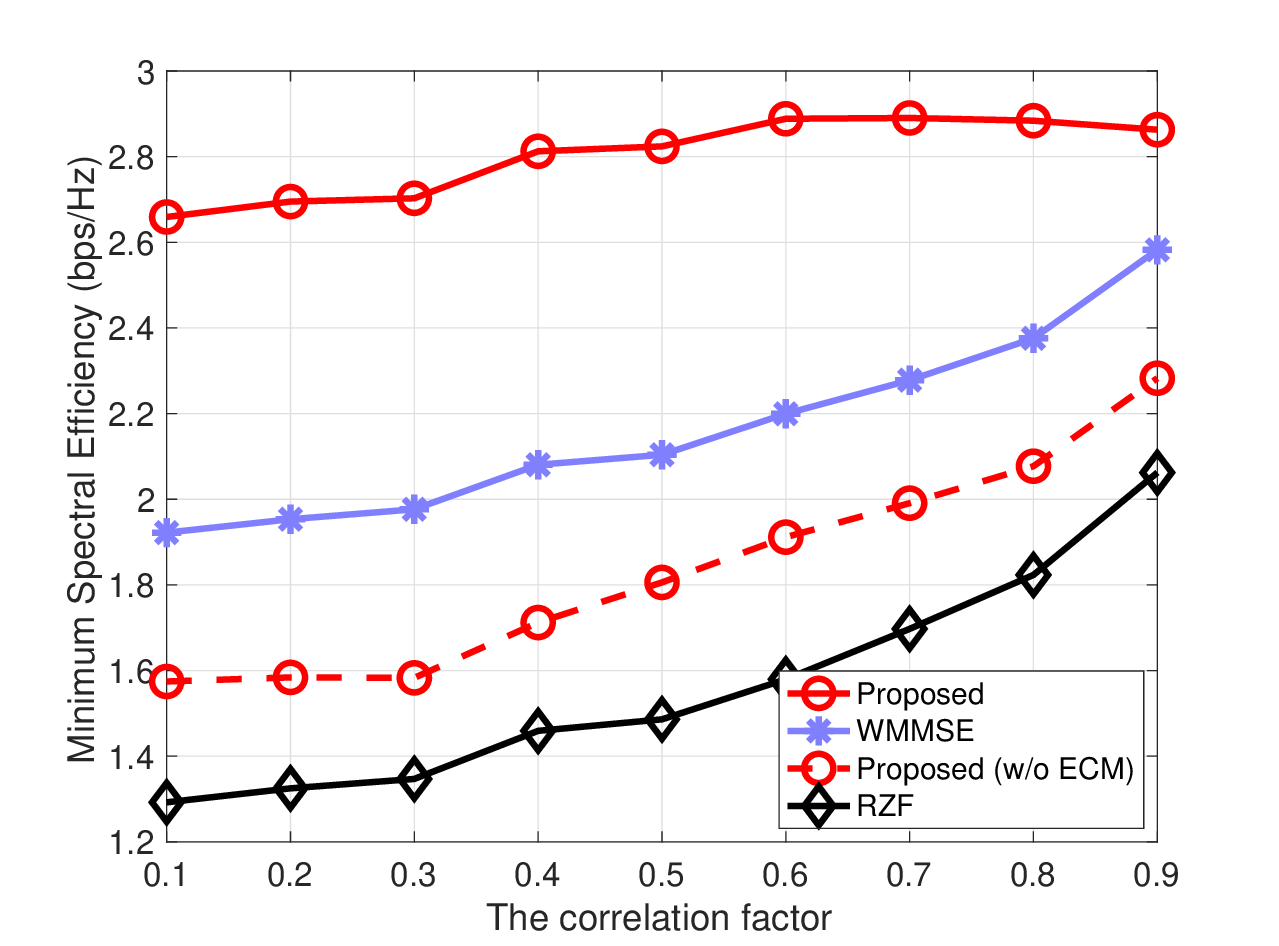}
    \caption{Comparison of the minimum spectral efficiency as a function of the path gain correlation factor $\eta_{k,l}$ at a fixed SNR of 30 dB.}
    \label{Fig:eta_sweep}
\end{figure}

Furthermore, we investigated the robustness of our framework to variations in the path gain correlation factor, $\eta_{k,\ell}$. While our primary simulations assume $\eta_{k,\ell} = 0.9$, this value can be lower in practical scenarios with higher environmental dynamics or channel aging effects. Although our framework targets fixed or moderate mobility, such environmental dynamics inevitably reduce the correlation between uplink and downlink channels. The simulation results presented in Fig.~\ref{Fig:eta_sweep} validate the framework's performance under such conditions. As shown in the figure, the performance of all methods degrades as $\eta_{k,\ell}$ decreases, reflecting the increased channel uncertainty. However, our proposed method with ECM maintains significantly more robust performance, exhibiting a much more graceful degradation compared to the case without ECM. This result underscores that our ECM approximation in \eqref{ECM_reciprocity}, which explicitly incorporates this uncertainty, is not only valid but becomes increasingly crucial for maintaining reliable performance in challenging environments with low channel reciprocity.

\subsection{Latency}

\begin{figure}[t]
 \renewcommand{\figurename}{Fig.}
    \centering
    \includegraphics[width=7cm]{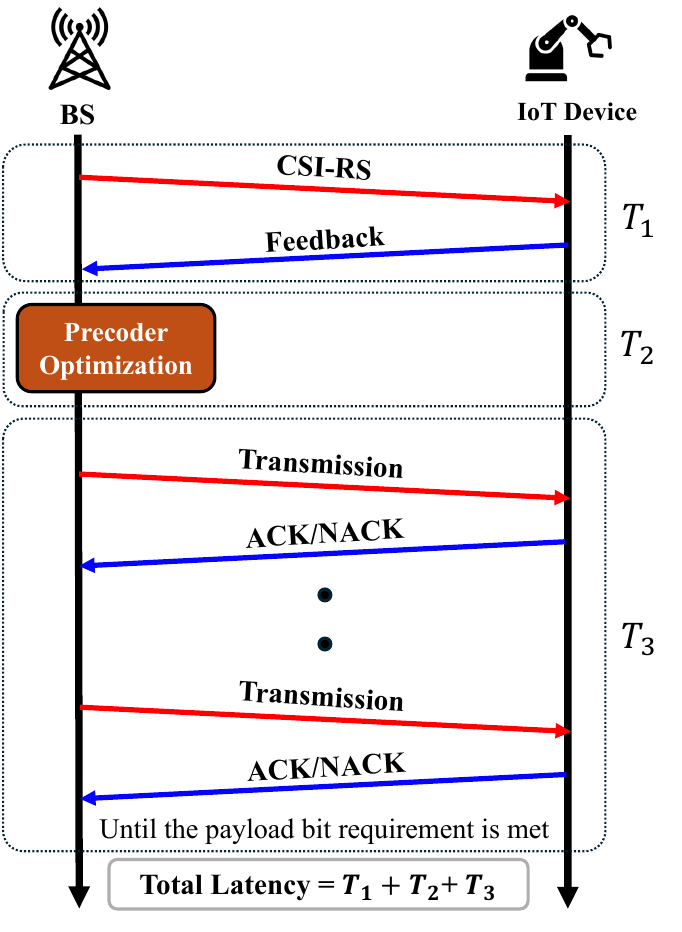}
    \caption{Transmission process in terms of latency.}
    \label{Fig:latency}
\end{figure}

We now evaluate the latency performance to provide a more rigorous demonstration of the proposed framework. 
To analyze the communication latency, we particularly focus on the following three processes, which constitute the major components of latency in practical 5G-NR systems. 

\begin{figure*}
\centering
\subfigure[Payload size = 25,000 bits (3.125 KB)]{\includegraphics[width=0.49\textwidth]{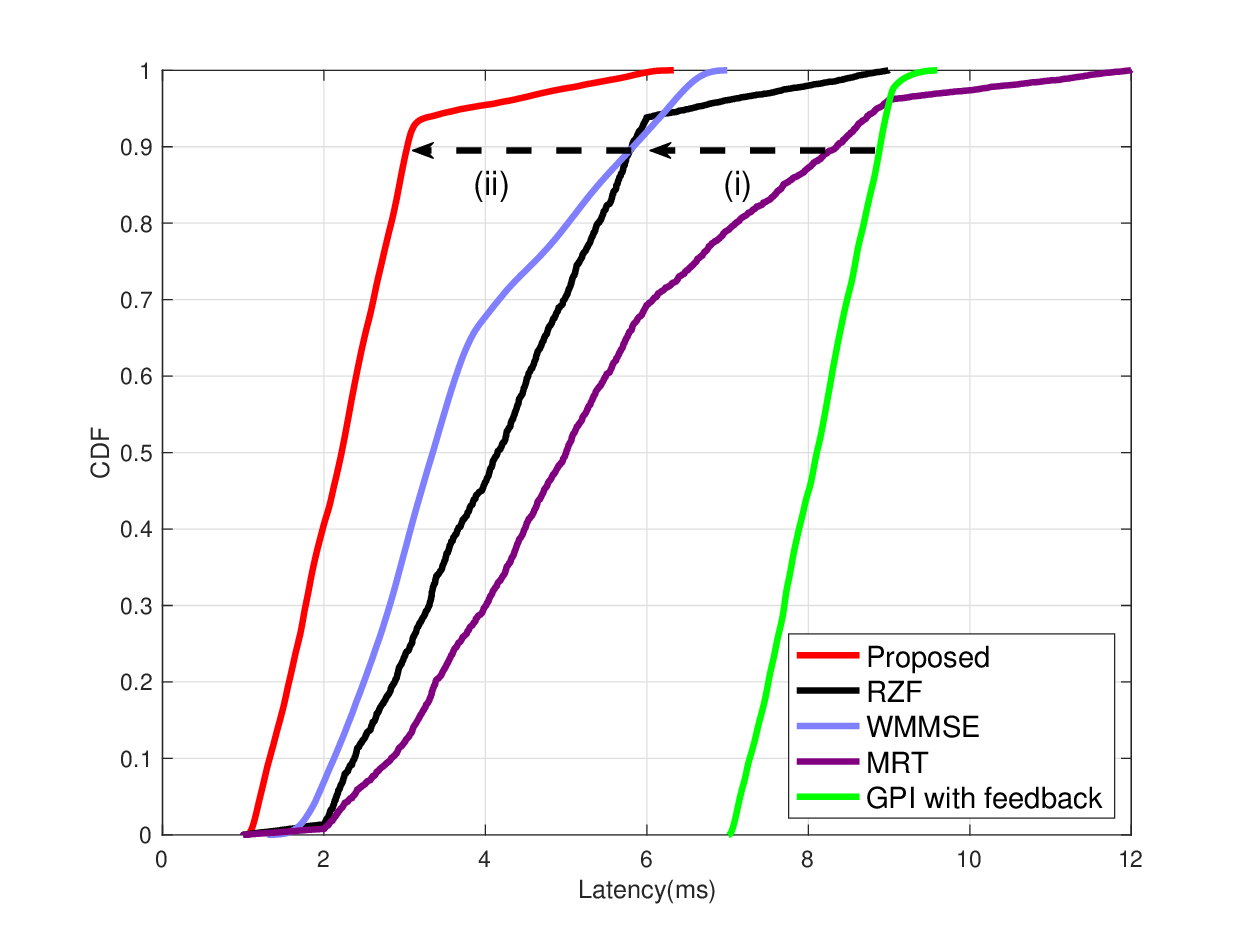}}\hfill
\subfigure[Payload size = 50,000 bits (6.25 KB)]{\includegraphics[width=0.49\textwidth]{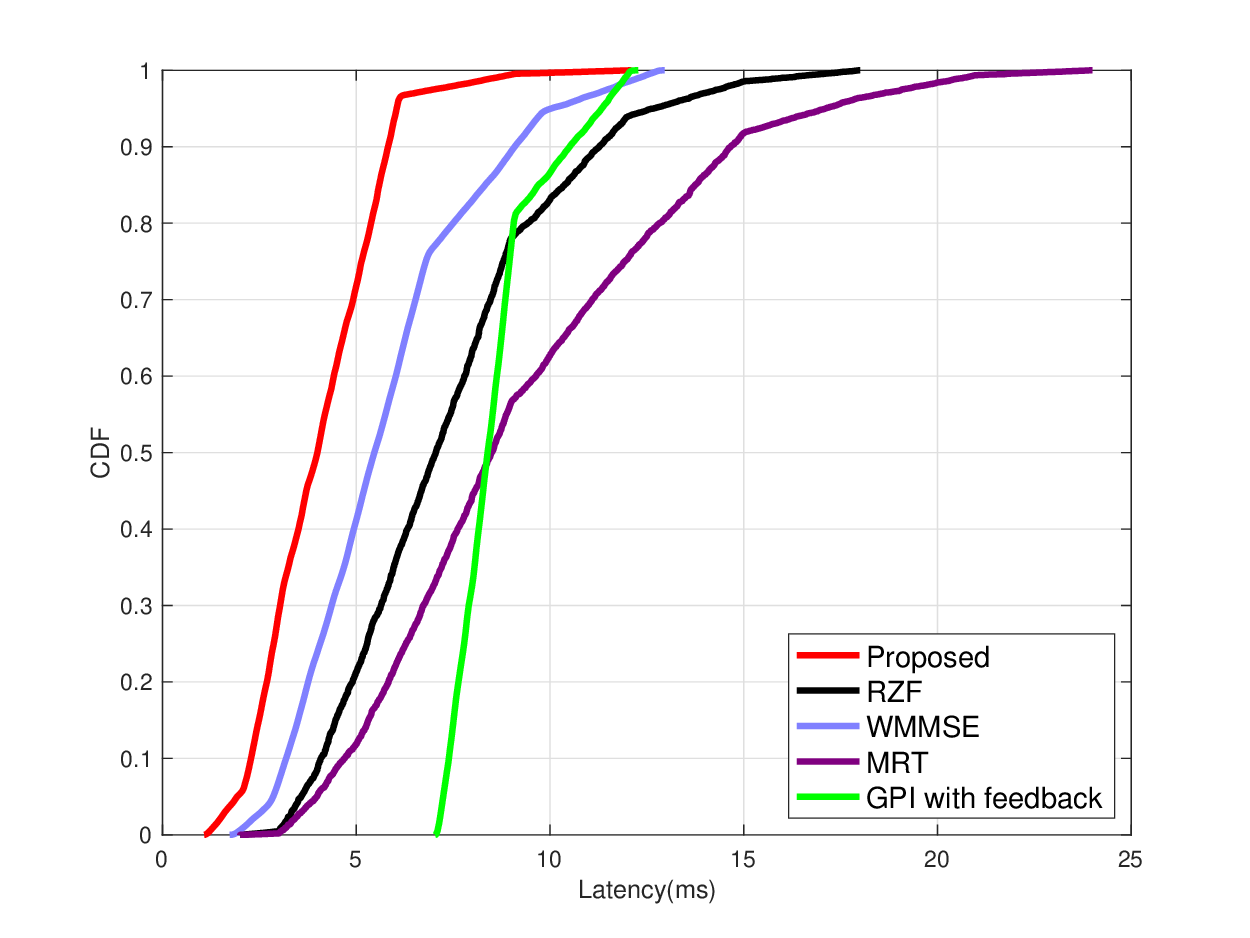}}
\caption{Comparison of the conventional feedback based transmission and the proposed uplink SRS based transmission, focusing on latency performance. The cumulative distribution function (CDF) of latency is plotted for payload sizes of 25,000 and 50,000 bits.}
\label{Fig:latency_cdf}
\end{figure*}

\begin{enumerate}
    \item \textbf{Downlink CSI acquisition}: This process counts the latency in acquiring downlink CSI. To be specific, following the conventional approach in 5G-NR, the BS first sends CSI-RS to the scheduled device. Subsequently, the device computes the CSI and sends the CSI feedback containing RI, PMI, and CQI to the BS \cite{3GPP}. Typically, in FDD MIMO, the entire process is known to take approximately $6-10$ ms. We denote the latency corresponding to the downlink CSI acquisition as $T_1 = 6$ ms.
    
    \item \textbf{Precoder optimization}: Given the downlink CSI, the BS runs the optimization algorithm to design the precoder. The latency of precoder optimization, denoted as $T_2$, is counted in this process. However, accurately measuring the actual latency associated with precoder optimization is very challenging. This is because it heavily depends on the implementation details of the precoder design algorithms, which are beyond the scope of this paper. Further, while \cite{3GPP} specifies timing requirements for device processing capabilities, the BS processing time requirements are not explicitly defined, as they are left to vendor implementation. 
    Incorporating practical BS computational constraints, we assume that the maximum allowable latency for FPGA-implemented precoder optimization, denoted as $T_2^{\sf max}$, is $1$ ms. 
    This assumption is supported by state-of-the-art FPGA implementations such as \cite{moon:tcas1:23}, where precoding optimization for MU-MIMO systems achieved a processing latency of $0.6$ ms using optimized hardware architectures. 
    Then, using WMMSE as a baseline, which consumes $1.94$ CPU time, we denote this baseline computation time as $T_2^{\sf max}$. Since the proposed method requires only $11.6\%$ of the WMMSE computation time, its relative processing time can be expressed as $T_2^{\sf max} \times 0.116$. The relative processing times of other baseline precoding methods, including MRT and RZF, are measured and normalized in the same manner. This estimation is motivated by the fundamental difference in computational complexity between the algorithms. Specifically, while the WMMSE algorithm entails a worst-case complexity of $\mathcal{O}((KN)^{3.5})$, the proposed GPI-based method reduces this to $\mathcal{O}(KN^3)$, implying a lower computational burden. While the actual latency in hardware implementations such as FPGAs may differ, we estimate the precoding latency by applying the computation time ratio to the baseline WMMSE latency $T_2^{\mathsf{max}}$.
    \item \textbf{Data transmission}: Using the precoder designed in the precoder optimization process, the BS sends the information symbols to the devices. 
    Assuming hybrid automatic repeat request (HARQ) with incremental redundancy (IR), the accumulated mutual information achieved in device $k$ during the $T$-th HARQ round is expressed as \cite{cerna:twc:24}
    \begin{align}
        I_k^{\sf acc.}{[T]} = \sum_{t = 1}^{T} \left\{ C_k[t] + R_{p,k}[t] \right\},
    \end{align}
    where $C_k[t]$ and $R_{p,k}[t]$ represent the achieved common rate portion and the private rate of device $k$ at the $t$-th HARQ round, respectively. 
    The total HARQ round $T^*$ is determined as
    \begin{align} \label{eq:harq_round}
        T^* = \min\{T | I_k^{\sf acc.}{[T]} \ge {\text{Payload}}_k \},
    \end{align}
    which represents the minimum number of HARQ rounds required for the accumulated mutual information to be greater than or equal to the given payload. In each HARQ round, the device sends an acknowledgement (ACK) or a negative-acknowledgement (NACK) signal back to the BS based on the decoding result. Consequently, the total latency in this process is computed as $T_3 = T^*\times T_3^{\sf ind.}$, where 
     $T_3^{\sf ind.}$ represents the latency of each HARQ round, including transmission air time, decoding, and feedback processing. Based on practical system parameters in \cite{3GPP}, we assume $T_3^{\sf ind.}$ to be in the range of 1.5 ms to 2.5 ms, which accounts for one slot transmission ($0.5$ ms with $30$ kHz subcarrier spacing), IoT device processing time (0.5 - 1.5 ms), and HARQ feedback processing ($0.5$ ms).
\end{enumerate}

The whole process is illustrated in Fig.~\ref{Fig:latency}. Since the proposed framework does not rely on any direct CSI acquisition process, we assume $T_1 = 0$ for the proposed framework. 
In contrast, for schemes using conventional downlink CSI acquisition processes, we assume perfect downlink CSI is obtained with $T_1 = 6$ ms. Notably, by reducing the feedback latency $T_1$ to zero, the proposed framework significantly minimizes the time interval between channel estimation and data transmission. This reduction ensures that the transmission process is completed well within the channel coherence time, thereby validating the assumption that channel parameters remain stable compared to conventional schemes.

The system operates with a fixed SNR of $30$ dB and a transmission bandwidth of $20$ MHz. 
Given the precoding method and bandwidth, we calculate the total number of HARQ rounds, and $T_3$ is subsequently determined as \eqref{eq:harq_round}. The total latency is then calculated as $T_1+T_2+T_3$. 
While this evaluation framework provides approximations rather than exact measurements, it enables fair comparison of latency performance across different precoding methods. 
For the payload size, we consider 25,000 bits (3.125 KB) and 50,000 bits (6.25 KB). 
We plot the latency cumulative distribution function (CDF) in Fig.~\ref{Fig:latency_cdf}. 

As shown in Fig.~\ref{Fig:latency_cdf}, the proposed method achieves substantially lower latency across both payload scenarios. At the $90$th percentile of IoT devices, the proposed method reduces latency by approximately $5.85$ ms compared to conventional approaches when the payload size is 25,000 bits. 
This latency reduction comes from directly estimating downlink CSIT from the uplink channel, thereby eliminating the need for CSIT feedback. In contrast, the feedback-based method requires a minimum latency of 6 ms even with perfect CSIT, primarily due to the feedback process delay of $T_1 = 6$ ms. This latency gap remains significant even when the payload size increases to 50,000 bits, where spectral efficiency becomes the dominant factor.

Among the feedback-free methods, the proposed approach achieves the lowest latency by applying the proposed robust precoder design that combines RSMA and error covariance matrix approximation. This design enhances spectral efficiency, allowing devices to meet their payload requirements in fewer transmission rounds. 
Furthermore, despite these sophisticated components, our GPI-based precoder optimization requires only $11.6\%$ of WMMSE's computation time, as shown in Table~\ref{Table:computation}. 
These combined advantages result in approximately $2.79$ ms lower latency at the $90$-th percentile of IoT devices, compared to the WMMSE method when the payload size is 25,000 bits.

While feedback-free RZF and MRT offer low computational complexity, their spectral efficiency is inherently limited, as demonstrated in Fig.~\ref{Fig:minimum_rate}. Consequently, these methods require more HARQ rounds, increasing $T_3$. As a result, more latency is required, specifically $2.78$ ms for RZF and $5.32$ ms for MRT at the $90$th percentile, compared to the proposed method. This performance gap becomes more pronounced with a payload size of 50,000 bits, where their limited spectral efficiency leads to substantially higher latency due to increased HARQ rounds.

While reconstructing the downlink CSI from uplink pilots using 2D-NOMP algorithm is the initial step toward eliminating feedback, this approach on its own is insufficient, yielding only the limited latency reduction labeled (i) in Fig.~\ref{Fig:latency_cdf}-(a). The more significant latency reduction, labeled (ii), is achieved only by deploying our integrated framework to resolve the subsequent technical bottlenecks. Specifically, by robustly combining our ECM approximation with RSMA, the framework enhances spectral efficiency, which directly reduces the required number of transmission rounds. This illustrates that our primary contribution is not merely the removal of feedback, but the holistic design of a complete and robust precoding system that makes a feedback-free operation both viable and highly effective for low-latency communications.

\subsection{Energy Efficiency}
To evaluate the system's performance for battery-constrained IoT devices, we now analyze the energy efficiency, a metric that captures the trade-off between throughput and power expenditure. The energy efficiency is defined as the ratio of the system's minimum spectral efficiency to the total power consumption of the BS and all IoT devices. Specifically, the energy efficiency is formulated as
\begin{align}
    \xi = \frac{K\cdot \min_{k \in \mathcal{K}} (C_k + \bar{R}_{p,k})}{P_{\text{BS}} + \sum_{k=1}^{K}(P_{k,\text{feedback}}(\cdot) + P_{k,\text{ADC}}(\cdot))}, \label{eq:ee_metric}
\end{align}
where the numerator is the minimum spectral efficiency among IoT devices, $P_{\text{BS}}$ is the BS transmit power $P$, and the denominator includes the IoT device power consumed for data reception ($P_{k,\text{ADC}}$) and CSI feedback ($P_{k,\text{feedback}}$).

The power consumption models for the device components follow \cite{goldsmith:twc:05, choi:twc:22}, where the key circuit power components are defined as:
\begin{itemize}
    \item $P_{k,\text{feedback}}(\cdot) = 2P_{k,\text{ADC}}(\cdot) + P_{\text{LO}} + 2P_{k,\text{DAC}}(\cdot) + P_{\text{RF}}$
    \item $P_{k,\text{ADC}}(\cdot)= 3V_{dd}^2 L_{\text{min}}\frac{f_s}{2}10^{(0.1525 b_{\text{ADC}}-4.838)}$
    \item $P_{k,\text{DAC}}(\cdot) = 1.5 \times 10^{-5} \cdot 2^{b_{\text{DAC}}} + 9\times 10^{-12} f_s b_{\text{DAC}}$
\end{itemize}
The relevant parameters are detailed in Table~\ref{tab:ee_params}. For our proposed framework and other feedback-free schemes, the feedback power is set to zero, i.e., $P_{k,\text{feedback}}(\cdot) = 0$.

\begin{table}[t]
\centering
\caption{Parameters for Energy Efficiency Model}
\label{tab:ee_params}
\begin{tabular}{|c|c|c|}
\hline\hline
\textbf{Parameter} & \textbf{Definition} & \textbf{Value} \\ \hline
$P_{\text{LO}}$    & Local oscillator power & 22.5 mW \\ \hline
$P_{\text{RF}}$    & RF chain power & 31.6 mW \\ \hline
$L_{\text{min}}$   & Minimum channel length & 0.5 $\mu$m \\ \hline
$V_{\text{dd}}$    & Power supply voltage & 3 V \\ \hline
$f_s$              & Sampling rate & $10^8$ Hz \\ \hline\hline
\end{tabular}
\end{table}

\begin{figure}[t]
 \renewcommand{\figurename}{Fig.}
    \centering
    \includegraphics[width=\columnwidth]{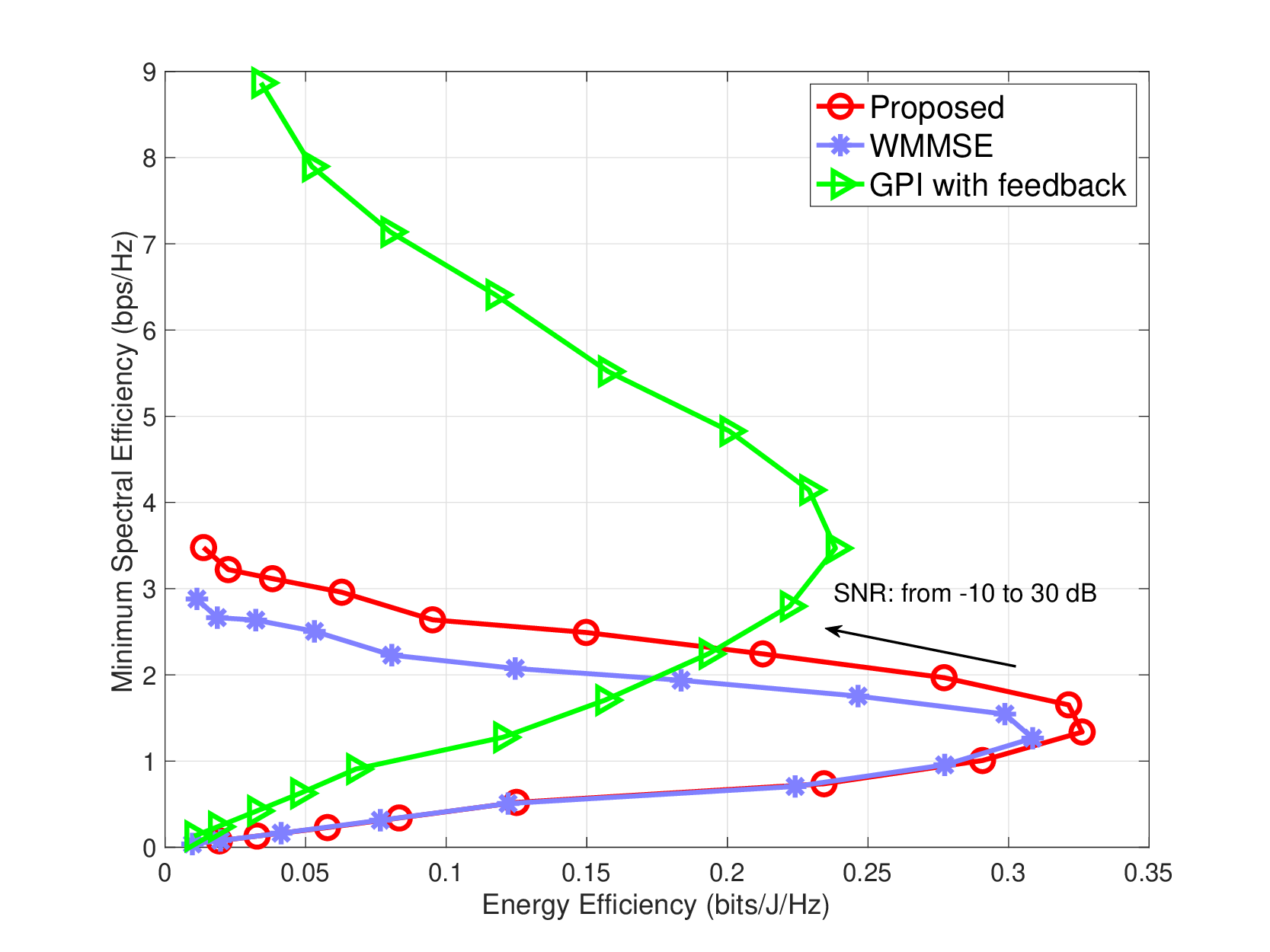}
    \caption{Energy efficiency versus SNR for different methods.}
    \label{Fig:energy_efficiency}
\end{figure}

For the numerical analysis, we evaluate the energy efficiency for SNR values ranging from $-10$ to $30$ dB, setting the quantization bits to $b_{\text{ADC}}=b_{\text{DAC}}=16$. The results are presented in Fig.~\ref{Fig:energy_efficiency}. As shown, the energy efficiency of all schemes decreases in the high SNR regime. This is because the BS transmit power ($P_{\text{BS}}=P$) grows more significantly than the spectral efficiency gain, becoming the dominant factor in the total power consumption.

The proposed method and the WMMSE benchmark show a clear energy efficiency advantage over the GPI with feedback method by eliminating the feedback power overhead. While the proposed method and WMMSE share the same total power consumption, our method achieves higher energy efficiency due to its superior minimum spectral efficiency, as established in Fig.~\ref{Fig:minimum_rate}. This spectral efficiency gain, when divided by the same power consumption, translates directly into an energy efficiency improvement. In contrast, despite assuming perfect CSIT, the GPI with feedback method's efficiency is significantly hampered by the additional power burden of the $P_{k,\text{feedback}}(\cdot)$ term, highlighting the practical benefits of our feedback-free framework for energy-constrained IoT networks.

\section{Conclusion}

In this paper, we presented a novel framework for enabling low-latency FDD MIMO transmission in IoT networks. By reconstructing downlink CSIT from uplink reference signals and addressing interference with RSMA, we significantly reduce communication overhead by eliminating CSIT feedback. To mitigate the effects of imperfect CSIT from the reconstruction, estimating the error covariance matrix further enhances robustness of the RSMA precoder. Additionally, the GPI-based approach efficiently computes the precoding vector, reducing computational complexity. Simulation results validate the performance of the robust precoder in maximizing the minimum spectral efficiency and the effectiveness of our approach in achieving low-latency. The proposed framework is developed under the assumption that channel parameters remain stable during the reconstruction process. Extending our framework to maintain robust performance in more dynamic and non-stationary environments, where these parameters may change over time, constitutes a promising direction for future research. Furthermore, investigating the impact of imperfect SIC in the finite blocklength regime remains an interesting topic for future work.

\bibliographystyle{IEEEtran}
\bibliography{ref_mmf_dark}

@string{globecom="Proc. IEEE Glob. Comm. Conf."}

@string{procieee="Proc. IEEE"}

@string{icc="Proc. IEEE Int. Conf. on Comm."}

@string{asilomar="Proc. of Asilomar Conf. on Sign., Syst. and Computers"}

@string{jsac="IEEE J. Sel. Areas Commun."}

@string{tvt="IEEE Trans. Veh. Technol."}

@string{tsp="IEEE Trans. Signal Process."}

@string{tcom="IEEE Trans. Commun."}

@string{twc="IEEE Trans. Wireless Commun."}

@string{tit="IEEE Trans. Inf. Theory"}

@string{ProcIEEE="Proc. of the IEEE"}

@string{tccn="IEEE Trans. Cognitive Comm. and Networking"}

@string{access="IEEE Access"}

@string{iotj="IEEE Internet of Things J."}

@string{commmag="IEEE Commun. Mag."}

@string{commlett="IEEE Commun. Lett."}

@string{wcl = "IEEE Wireless Commun. Lett."}

@string{survey="IEEE Commun. Surveys \& Tutorials"}

@ARTICLE{osse:commmag:14,
author={A. {Osseiran} and F. {Boccardi} and V. {Braun} and K. {Kusume} and P. {Marsch} and M. {Maternia} and O. {Queseth} and M. {Schellmann} and H. {Schotten} and H. {Taoka} and H. {Tullberg} and M. A. {Uusitalo} and B. {Timus} and M. {Fallgren}},
journal=commmag, 
title={Scenarios for 5{G} mobile and wireless communications: the vision of the {METIS} project}, 
year={2014},
volume={52},
number={5},
pages={26-35}
}

@ARTICLE{durisi:proc:16,
author={G. {Durisi} and T. {Koch} and P. {Popovski}},
journal={Proc. of IEEE}, 
title={Toward Massive, Ultrareliable, and Low-Latency Wireless Communication With Short Packets}, 
year={2016},
volume={104},
number={9},
pages={1711-1726}
}

@ARTICLE{kim:wcl:23,
  author={Kim, Doseon and Choi, Jinseok and Park, Jeonghun and Kim, Dong Ku},
  journal=wcl, 
  title={Max–Min Fairness Beamforming With Rate-Splitting Multiple Access: {Optimization} Without a Toolbox}, 
  year={2023},
  volume={12},
  number={2},
  pages={232-236},
  doi={10.1109/LWC.2022.3221526}}

@ARTICLE{poly:tit:10, 
author={Y. {Polyanskiy} and H. V. {Poor} and S. {Verdu}}, 
journal=tit, 
title={Channel Coding Rate in the Finite Blocklength Regime}, 
year={2010}, 
volume={56}, 
number={5}, 
pages={2307-2359}, 
month=may
}

@ARTICLE{ghanem:tcom:20,
author={W. R. {Ghanem} and V. {Jamali} and Y. {Sun} and R. {Schober}},
journal=tcom, 
title={Resource Allocation for Multi-User Downlink {MISO OFDMA-URLLC} Systems}, 
year={2020},
volume={},
number={},
pages={1-1},
month={}
}

@ARTICLE{schiessl:jsac:19,
author={S. {Schiessl} and J. {Gross} and M. {Skoglund} and G. {Caire}},
journal=jsac, 
title={Delay Performance of the Multiuser {MISO} Downlink Under Imperfect {CSI} and Finite-Length Coding}, 
year={2019},
volume={37},
number={4},
pages={765-779}
}

@ARTICLE{yang:tit:14,
author={W. {Yang} and G. {Durisi} and T. {Koch} and Y. {Polyanskiy}},
journal=tit, 
title={Quasi-Static Multiple-Antenna Fading Channels at Finite Blocklength}, 
year={2014},
volume={60},
number={7},
pages={4232-4265}
}

@ARTICLE{schiessl:tcom:18,
  author={Schiessl, Sebastian and Al-Zubaidy, Hussein and Skoglund, Mikael and Gross, James},
  journal=tcom, 
  title={Delay Performance of Wireless Communications With Imperfect {CSI} and Finite-Length Coding}, 
  year={2018},
  volume={66},
  number={12},
  pages={6527-6541},
  doi={10.1109/TCOMM.2018.2860000}}

@ARTICLE{schiessl:twc:20,
  author={Schiessl, Sebastian and Skoglund, Mikael and Gross, James},
  journal=twc, 
  title={{NOMA} in the Uplink: Delay Analysis With Imperfect {CSI} and Finite-Length Coding}, 
  year={2020},
  volume={19},
  number={6},
  pages={3879-3893},
  doi={10.1109/TWC.2020.2979114}}

@ARTICLE{choi:iotj:21,
  author={Choi, Jinseok and Park, Jeonghun},
  journal=iotj, 
  title={{MIMO} Design for Internet of Things: {Joint} Optimization of Spectral Efficiency and Error Probability in Finite Blocklength Regime}, 
  year={2021},
  volume={8},
  number={20},
  pages={15512-15521},
  doi={10.1109/JIOT.2021.3073239}}

@ARTICLE{kim:twc:22,
  author={Kim, Minsu and Park, Jeonghun and Lee, Jemin},
  journal=twc, 
  title={Precoding Design for Multi-User {MISO} Systems With Delay-Constrained and -Tolerant Users}, 
  year={2022},
  volume={21},
  number={7},
  pages={5090-5105},
  doi={10.1109/TWC.2021.3136872}}

@ARTICLE{FDD-delay,
  author={Lee, Byungju and Choi, Junil and Seol, Ji-Yun and Love, David J. and Shim, Byonghyo},
  journal=tcom, 
  title={Antenna Grouping Based Feedback Compression for {FDD}-Based Massive {MIMO} Systems}, 
  year={2015},
  volume={63},
  number={9},
  pages={3261-3274},
  doi={10.1109/TCOMM.2015.2460743}}

@ARTICLE{FDD-Low-Freq,
  author={Jeon, Jeongho and Lee, Gilwon and Ibrahim, Ahmad A.I. and Yuan, Jin and Xu, Gary and Cho, Joonyoung and Onggosanusi, Eko and Kim, Younsun and Lee, Juho and Zhang, Jianzhong Charlie},
  journal={IEEE Commun. Mag.}, 
  title={{MIMO} Evolution toward {6G}: Modular Massive {MIMO} in Low-Frequency Bands}, 
  year={2021},
  volume={59},
  number={11},
  pages={52-58},
  doi={10.1109/MCOM.211.2100164}}

@ARTICLE{FDD-coverage,
  author={Zeydan, Engin and Dedeoglu, Omer and Turk, Yekta},
  journal={IEEE Access}, 
  title={Experimental Evaluations of {TDD}-Based Massive {MIMO} Deployment for Mobile Network Operators}, 
  year={2020},
  volume={8},
  number={},
  pages={33202-33214},
  doi={10.1109/ACCESS.2020.2974277}}

@ARTICLE{JOMP,
  author={Rao, Xiongbin and Lau, Vincent K. N.},
  journal=tsp, 
  title={Distributed Compressive {CSIT} Estimation and Feedback for {FDD} Multi-User Massive {MIMO} Systems}, 
  year={2014},
  volume={62},
  number={12},
  pages={3261-3271},
  doi={10.1109/TSP.2014.2324991}}

@INPROCEEDINGS{Deeplearning-Ahmed,
  author={Alrabeiah, Muhammad and Alkhateeb, Ahmed},
  booktitle=asilomar, 
  title={Deep Learning for {TDD and FDD} Massive {MIMO}: Mapping Channels in Space and Frequency}, 
  year={2019},
  volume={},
  number={},
  pages={1465-1470},
  doi={10.1109/IEEECONF44664.2019.9048929}}

@ARTICLE{guo:tcom:22,
  author={Guo, Jiajia and Wen, Chao-Kai and Jin, Shi and Li, Geoffrey Ye},
  journal=tcom, 
  title={Overview of Deep Learning-Based {CSI} Feedback in Massive {MIMO} Systems}, 
  year={2022},
  volume={70},
  number={12},
  pages={8017-8045}
}

@inproceedings{Dina,
  title={Eliminating channel feedback in next-generation cellular networks},
  author={Vasisht, Deepak and Kumar, Swarun and Rahul, Hariharan and Katabi, Dina},
  booktitle={Proc. of the ACM SIGCOMM Conf.},
  pages={398--411},
  year={2016}
}

@ARTICLE{T.Choi:2021,
  author={Choi, Thomas and Rottenberg, François and Gómez-Ponce, Jorge and Ramesh, Akshay and Luo, Peng and Zhang, Charlie Jianzhong and Molisch, Andreas F.},
  journal=twc, 
  title={Experimental Investigation of Frequency Domain Channel Extrapolation in Massive {MIMO} Systems for Zero-Feedback {FDD}}, 
  year={2021},
  volume={20},
  number={1},
  pages={710-725},
  doi={10.1109/TWC.2020.3028161}}

@ARTICLE{Rottenberg:2020,
author = {F. Rottenberg and T. Choi and P. Luo and C. J. Zhang and A. F. Molisch},
title  = {Performance Analysis of Channel Extrapolation in {FDD} Massive {MIMO} Systems},
journal = twc,
volume = "19",
number = "4",
pages = {2728-2741},
month = {Apr.},
year      = {2020}
}

@ARTICLE{Zhong:2020,
author = {Z. Zhong and L. Fan and S. Ge},
title = {{FDD} Massive {MIMO} Uplink and Downlink Channel Reciprocity Properties: Full or Partial Reciprocity?},
journal = globecom,
month = {Dec.},
year      = {2020}
}

@ARTICLE{Deokhwan,
  author={Han, Deokhwan and Park, Jeonghun and Lee, Namyoon},
  journal=twc, 
  title={{FDD} Massive {MIMO} Without {CSI} Feedback}, 
  year={2023},
  volume={},
  number={},
  pages={1-1},
  doi={10.1109/TWC.2023.3319486}}

@ARTICLE{Han:2019,
author = {Y. Han and T. -H. Hsu and C. -K. Wen and K. -K. Wong and S. Jin},
title = {Efficient Downlink Channel Reconstruction for {FDD} Multi-Antenna Systems},
journal = twc,
volume = "18",
number = "6",
pages = {3161-3176},
month = {June},
year      = {2019}
}

@ARTICLE{han:tcom:19,
  author={Han, Yu and Liu, Qi and Wen, Chao-Kai and Jin, Shi and Wong, Kai-Kit},
  journal=tcom, 
  title={{FDD} Massive {MIMO} Based on Efficient Downlink Channel Reconstruction}, 
  year={2019},
  volume={67},
  number={6},
  pages={4020-4034},
  doi={10.1109/TCOMM.2019.2900625}}

@ARTICLE{zhang:twc:18,
  author={Zhang, Xing and Zhong, Lin and Sabharwal, Ashutosh},
  journal=twc, 
  title={Directional Training for {FDD} Massive {MIMO}}, 
  year={2018},
  volume={17},
  number={8},
  pages={5183-5197},
  doi={10.1109/TWC.2018.2838600}}

@ARTICLE{NOMP,
author = {B. Mamandipoor and D. Ramasamy and U. Madhow},
title  = "Newtonized Orthogonal Matching Pursuit: Frequency Estimation Over the Continuum",
journal = tsp,
volume = "64",
number = "19",
pages = {5066-5081},
month = {Oct.},
year      = {2016}
}

@ARTICLE{Park:2023,
author = {J. Park and J. Choi and N. Lee and W. Shin and H. V. Poor},
title  = "Rate-Splitting Multiple Access for Downlink {MIMO}: A Generalized Power Iteration Approach",
journal = twc,
volume = "22",
number = "3",
month = "Mar.",
year  = {2023},
pages = "1588-1603"
}

@ARTICLE{RSMA-ten-promising,
  author={Park, Jeonghun and Lee, Byungju and Choi, Jinseok and Lee, Hoon and Lee, Namyoon and Park, Seok-Hwan and Lee, Kyoung-Jae and Choi, Junil and Chae, Sung Ho and Jeon, Sang-Woon and Kwak, Kyung Sup and Clerckx, Bruno and Shin, Wonjae},
  journal={IEEE Network}, 
  title={Rate-Splitting Multiple Access for {6G} Networks: {Ten} Promising Scenarios and Applications}, 
  year={2023},
  volume={},
  number={},
  pages={1-1},
  doi={10.1109/MNET.2023.3321518}}

@ARTICLE{xu:tvt:22,
  author={Xu, Yunnuo and Mao, Yijie and Dizdar, Onur and Clerckx, Bruno},
  journal=tvt, 
  title={Rate-Splitting Multiple Access With Finite Blocklength for Short-Packet and Low-Latency Downlink Communications}, 
  year={2022},
  volume={71},
  number={11},
  pages={12333-12337},
  doi={10.1109/TVT.2022.3191085}}

@ARTICLE{wang:jsac:23,
  author={Wang, Yuan and Wong, Vincent W. S. and Wang, Jiaheng},
  journal=jsac, 
  title={Flexible Rate-Splitting Multiple Access With Finite Blocklength}, 
  year={2023},
  volume={41},
  number={5},
  pages={1398-1412},
  doi={10.1109/JSAC.2023.3240783}}

@ARTICLE{mao:tcom:21,
  author={Mao, Yijie and Piovano, Enrico and Clerckx, Bruno},
  journal=tcom, 
  title={Rate-Splitting Multiple Access for Overloaded Cellular Internet of Things}, 
  year={2021},
  volume={69},
  number={7},
  pages={4504-4519},
  doi={10.1109/TCOMM.2021.3067642}}

@ARTICLE{zhang:iotj:20,
  author={Zhang, Liang and Ansari, Nirwan},
  journal={IEEE Internet of Things J.}, 
  title={Latency-Aware {IoT} Service Provisioning in {UAV}-Aided Mobile-Edge Computing Networks}, 
  year={2020},
  volume={7},
  number={10},
  pages={10573-10580},
  doi={10.1109/JIOT.2020.3005117}}

@ARTICLE{ghosh:tccn:24,
  author={Ghosh, Sutanu and Singh, Keshav and Jung, Haejoon and Li, Chih-Peng and Duong, Trung Q.},
  journal=tccn, 
  title={On the Performance of Rate Splitting Multiple Access for {ISAC} in Device-to-Multi-Device {IoT} Communications}, 
  year={2024},
  volume={},
  number={},
  pages={1-1},
  doi={10.1109/TCCN.2024.3438359}}

@ARTICLE{qiu:iotj:24,
  author={Qiu, Bin and Cheng, Wenchi and Zhang, Wei},
  journal={IEEE Internet of Things J.}, 
  title={Joint Information and Jamming Beamforming for Securing {IoT} Networks With Ratesplitting}, 
  year={2024},
  volume={11},
  number={4},
  pages={6338-6351},
  doi={10.1109/JIOT.2023.3313870}}

@ARTICLE{caire:tsp:17,
  author={Haghighatshoar, Saeid and Caire, Giuseppe},
  journal=tsp, 
  title={Massive {MIMO} Channel Subspace Estimation From Low-Dimensional Projections}, 
  year={2017},
  volume={65},
  number={2},
  pages={303-318},
  doi={10.1109/TSP.2016.2616336}}

@ARTICLE{yin:jsac:2013,
  author={Yin, Haifan and Gesbert, David and Filippou, Miltiades and Liu, Yingzhuang},
  journal=jsac, 
  title={A Coordinated Approach to Channel Estimation in Large-Scale Multiple-Antenna Systems}, 
  year={2013},
  volume={31},
  number={2},
  pages={264-273},
  doi={10.1109/JSAC.2013.130214}}

@ARTICLE{duarte:11,
title = {Spectral compressive sensing},
journal = {Applied and Computational Harmonic Analysis},
volume = {35},
number = {1},
pages = {111-129},
year = {2013},
issn = {1063-5203},
doi = {https://doi.org/10.1016/j.acha.2012.08.003},
url = {https://www.sciencedirect.com/science/article/pii/S1063520312001315},
author = {Marco F. Duarte and Richard G. Baraniuk}
}

@ARTICLE{tropp:tit:07,
  author={Tropp, Joel A. and Gilbert, Anna C.},
  journal=tit, 
  title={Signal Recovery From Random Measurements Via Orthogonal Matching Pursuit}, 
  year={2007},
  volume={53},
  number={12},
  pages={4655-4666},
  doi={10.1109/TIT.2007.909108}}

@ARTICLE{kim:arxiv:24,
  author={Kim, Namhyun and Roberts, Ian P. and Park, Jeonghun},
  journal=twc, 
  title={Splitting Messages in the Dark-Rate-Splitting Multiple Access for {FDD} Massive {MIMO} Without {CSI} Feedback}, 
  year={2025},
  volume={24},
  number={4},
  pages={3320-3332},
  doi={10.1109/TWC.2025.3529945}}

@ARTICLE{park:twc:16,
  author={Park, Jeonghun and Lee, Namyoon and Andrews, Jeffrey G. and Heath, Robert W.},
  journal=twc, 
  title={On the Optimal Feedback Rate in Interference-Limited Multi-Antenna Cellular Systems}, 
  year={2016},
  volume={15},
  number={8},
  pages={5748-5762},
  doi={10.1109/TWC.2016.2569089}}

@ARTICLE{xie:twc:18,
  author={Xie, Hongxiang and Gao, Feifei and Jin, Shi and Fang, Jun and Liang, Ying-Chang},
  journal=twc, 
  title={Channel Estimation for {TDD/FDD} Massive {MIMO} Systems With Channel Covariance Computing}, 
  year={2018},
  volume={17},
  number={6},
  pages={4206-4218},
  doi={10.1109/TWC.2018.2821667}}

@ARTICLE{joudeh:tsp:16,
  author={Joudeh, Hamdi and Clerckx, Bruno},
  journal=tsp, 
  title={Robust Transmission in Downlink Multiuser {MISO} Systems: {A} Rate-Splitting Approach}, 
  year={2016},
  volume={64},
  number={23},
  pages={6227-6242},
  doi={10.1109/TSP.2016.2591501}}

@ARTICLE{zhu:iotj:24,
  author={Zhu, Jianyue and Jin, Haijia and He, Yutong and Fang, Fang and Huang, Wei and Zhang, Zhizhong},
  journal=iotj, 
  title={Joint Optimization of User Scheduling, Rate Allocation, and Beamforming for RSMA Finite Blocklength Transmission}, 
  year={2024},
  volume={11},
  number={17},
  pages={27904-27915},
  doi={10.1109/JIOT.2024.3420099}}

@ARTICLE{kim:proc:19,
  author={Kim, Kwang Soon and Kim, Dong Ku and Chae, Chan-Byoung and Choi, Sunghyun and Ko, Young-Chai and Kim, Jonghyun and Lim, Yeon-Geun and Yang, Minho and Kim, Sundo and Lim, Byungju and Lee, Kwanghoon and Ryu, Kyung Lin},
  journal=ProcIEEE, 
  title={Ultrareliable and Low-Latency Communication Techniques for Tactile Internet Services}, 
  year={2019},
  volume={107},
  number={2},
  pages={376-393},
  doi={10.1109/JPROC.2018.2868995}}

@ARTICLE{wmmse,
  author={Christensen, Søren Skovgaard and Agarwal, Rajiv and De Carvalho, Elisabeth and Cioffi, John M.},
  journal=twc, 
  title={Weighted sum-rate maximization using weighted MMSE for MIMO-BC beamforming design}, 
  year={2008},
  volume={7},
  number={12},
  pages={4792-4799},
  doi={10.1109/T-WC.2008.070851}}

@ARTICLE{song:iotj:24,
  author={Song, Shaoqian and Hu, Fengye and Ling, Zhuang and Li, Zhuofei and Jin, Chi},
  journal=iotj, 
  title={Max—Min Fairness of CR-RSMA-Based UAV Relay-Assisted Emergency Communication Network With Limited User Energy}, 
  year={2024},
  volume={11},
  number={13},
  pages={23998-24012},
  doi={10.1109/JIOT.2024.3390212}}

@ARTICLE{hu:tcom:23,
  author={Hu, Guojie and Li, Zan and Si, Jiangbo and Xu, Kui and Xu, Donghui and Cai, Yunlong and Al-Dhahir, Naofal},
  journal=tcom, 
  title={Maxmin Fairness for {UAV}-Enabled Proactive Eavesdropping With Jamming Over Distributed Transmit Beamforming-Based Suspicious Communications}, 
  year={2023},
  volume={71},
  number={3},
  pages={1595-1614},
  doi={10.1109/TCOMM.2023.3237258}}

@techreport{3GPP,
  title = {{3GPP TS} 38.214: {NR}; {Physical} layer procedures for data},
  author={3GPP},
  institution = {3rd Generation Partnership Project},
  type = {Technical Specification},
  version = {17.5.0},
  year = {2023},
  month = {March}
}

@techreport{3GPP_iot,
  title = {{3GPP TS} 38.101-1: {NR}; User Equipment (UE) radio transmission and reception; Part 1: Range 1 Standalone},
  author={3GPP},
  institution = {3rd Generation Partnership Project},
  type = {Technical Specification},
  version = {17.5.0},
  year = {2022},
  month = {May}
}

@techreport{3GPP_fim,
  title = {{3GPP TS} 38.802: Study on new radio access technology: Physical layer aspects},
  author={3GPP},
  institution = {3rd Generation Partnership Project},
  type = {Technical Specification},
  version = {14.2.0},
  year = {2017},
  month = {September}
}

@ARTICLE{vaezi:coms:2022,
  author={Vaezi, Mojtaba and Azari, Amin and Khosravirad, Saeed R. and Shirvanimoghaddam, Mahyar and Azari, M. Mahdi and Chasaki, Danai and Popovski, Petar},
  journal=survey, 
  title={Cellular, Wide-Area, and Non-Terrestrial IoT: A Survey on {5G} Advances and the Road Toward 6G}, 
  year={2022},
  volume={24},
  number={2},
  pages={1117-1174},
  doi={10.1109/COMST.2022.3151028}}

@ARTICLE{varsier:comm:2021,
  author={Varsier, Nadège and Dufrène, Louis-Adrien and Dumay, Marion and Lampin, Quentin and Schwoerer, Jean},
  journal=commmag, 
  title={A {5G} New Radio for Balanced and Mixed IoT Use Cases: Challenges and Key Enablers in {FR1} Band}, 
  year={2021},
  volume={59},
  number={4},
  pages={82-87},
  doi={10.1109/MCOM.001.2000660}}

@ARTICLE{kim:twc:24,
  author={Kim, Jungyeon and Choi, Jinseok and Park, Jeonghun and Alkhateeb, Ahmed and Lee, Namyoon},
  journal=twc, 
  title={{FDD} Massive {MIMO}: {How} to Optimally Combine {UL} Pilot and Limited {DL CSI} Feedback?}, 
  year={2024},
  volume={},
  number={},
  pages={1-1},
  doi={10.1109/TWC.2024.3496933}}

@ARTICLE{cerna:twc:24,
  author={Cerna Loli, Rafael and Dizdar, Onur and Clerckx, Bruno and Popovski, Petar},
  journal=twc, 
  title={Hybrid Automatic Repeat Request for Downlink Rate-Splitting Multiple Access}, 
  year={2024},
  volume={23},
  number={10},
  pages={15261-15275},
  doi={10.1109/TWC.2024.3427708}}

@MISC{cvx,
author = {Michael Grant and Stephen Boyd},
title = {{CVX}: Matlab Software for Disciplined Convex Programming, version 2.1},
howpublished = {\url{http://cvxr.com/cvx}},
month = mar,
year = 2014}

@ARTICLE{moon:tcas1:23,
  author={Moon, Seungsik and Lee, Namyoon and Lee, Youngjoo},
  journal={IEEE Trans. Circuits Syst. I}, 
  title={A Scalable Precoding Processor for Large-Scale {MU-MIMO} Systems}, 
  year={2023},
  volume={70},
  number={7},
  pages={3029-3039},
  doi={10.1109/TCSI.2023.3268314}}

@ARTICLE{caire:tit:13,
  author={Adhikary, Ansuman and Nam, Junyoung and Ahn, Jae-Young and Caire, Giuseppe},
  journal=tit, 
  title={Joint Spatial Division and Multiplexing—The Large-Scale Array Regime}, 
  year={2013},
  volume={59},
  number={10},
  pages={6441-6463},
  doi={10.1109/TIT.2013.2269476}}

@ARTICLE{shi:WCL:18,
  author={Wen, Chao-Kai and Shih, Wan-Ting and Jin, Shi},
  journal=commlett, 
  title={Deep Learning for Massive {MIMO CSI} Feedback}, 
  year={2018},
  volume={7},
  number={5},
  pages={748-751},
  doi={10.1109/LWC.2018.2818160}}

@ARTICLE{ganti:tit:00,
  author={Ganti, A. and Lapidoth, A. and Telatar, I.E.},
  journal=tit, 
  title={Mismatched decoding revisited: general alphabets, channels with memory, and the wide-band limit}, 
  year={2000},
  volume={46},
  number={7},
  pages={2315-2328},
  doi={10.1109/18.887846}}

@ARTICLE{goldsmith:twc:05,
  author={Shuguang Cui and Goldsmith, A.J. and Bahai, A.},
  journal=twc, 
  title={Energy-constrained modulation optimization}, 
  year={2005},
  volume={4},
  number={5},
  pages={2349-2360},
  doi={10.1109/TWC.2005.853882}}

@ARTICLE{choi:twc:22,
  author={Choi, Jinseok and Park, Jeonghun and Lee, Namyoon},
  journal=twc, 
  title={Energy Efficiency Maximization Precoding for Quantized Massive {MIMO} Systems}, 
  year={2022},
  volume={21},
  number={9},
  pages={6803-6817},
  doi={10.1109/TWC.2022.3152491}}

@ARTICLE{ma:twc:25,
  author={Ma, Yifan and He, Hengtao and Song, Shenghui and Zhang, Jun and Letaief, Khaled B.},
  journal={IEEE Transactions on Wireless Communications}, 
  title={Low-Complexity CSI Feedback for FDD Massive MIMO Systems via Learning to Optimize}, 
  year={2025},
  volume={24},
  number={4},
  pages={3483-3498},
  doi={10.1109/TWC.2025.3531447}}

@ARTICLE{vu:iotj:25,
  author={Vu, Thai-Hoc and Le, Anh-Tu and Hoang Tu, Ngo and Nguyen, Tan N. and Voznak, Miroslav},
  journal={IEEE Internet of Things Journal}, 
  title={On Performance of IoT Networks With Coordinated NOMA Transmission: Covert Monitoring and Information Decoding}, 
  year={2025},
  volume={12},
  number={22},
  pages={48069-48084},
  doi={10.1109/JIOT.2025.3605276}}

@ARTICLE{truong:jsac:25,
  author={Phung Truong, Thanh and My Tuyen Nguyen, Thi and Vi Nguyen, The and Dao, Nhu-Ngoc and Cho, Sungrae},
  journal={IEEE Journal on Selected Areas in Communications}, 
  title={Energy Efficiency in RSMA-Enhanced Active RIS-Aided Quantized Downlink Systems}, 
  year={2025},
  volume={43},
  number={3},
  pages={834-850},
  doi={10.1109/JSAC.2025.3531522}}

@ARTICLE{kim:tvt:25,
  author={Kim, Jeongbin and Jeong, Seongah and Yoo, Seonghoon and Son, Woong and Kang, Joonhyuk},
  journal={IEEE Transactions on Vehicular Technology}, 
  title={Rate-Splitting Multiple Access for Hierarchical HAP-LAP Networks Under Limited Fronthaul}, 
  year={2025},
  volume={74},
  number={8},
  pages={13173-13178},
  doi={10.1109/TVT.2025.3552053}}

@INPROCEEDINGS{sun:icc:18,
  author={Sun, Haijian and Zhou, Fuhui and Zhang, Zekun},
  booktitle={2018 IEEE International Conference on Communications (ICC)}, 
  title={Robust Beamforming Design in a NOMA Cognitive Radio Network Relying on SWIPT}, 
  year={2018},
  volume={},
  number={},
  pages={1-6},
  doi={10.1109/ICC.2018.8422527}}

@ARTICLE{le:tgcn:17,
  author={Le, Tuan Anh and Vien, Quoc-Tuan and Nguyen, Huan X. and Ng, Derrick Wing Kwan and Schober, Robert},
  journal={IEEE Transactions on Green Communications and Networking}, 
  title={Robust Chance-Constrained Optimization for Power-Efficient and Secure SWIPT Systems}, 
  year={2017},
  volume={1},
  number={3},
  pages={333-346},
  doi={10.1109/TGCN.2017.2706063}}


\end{document}